\documentclass[conference]{IEEEtran}

\IEEEoverridecommandlockouts

\usepackage{epsfig,amsmath,amsfonts,multirow,makecell,caption,soul,csquotes,color,wrapfig,subcaption,mathtools,bm,spverbatim,booktabs,xcolor,amsthm,tcolorbox,wrapfig,enumitem}
\usepackage[e]{esvect}
\usepackage{graphics}
\usepackage{marvosym,listings,etoolbox}
\usepackage{adjustbox}
\usepackage{cite}
\usepackage{subcaption}
\DeclareCaptionSubType * [alph]{table}
\usepackage{multicol}
\usepackage{lipsum}

\makeatletter
\renewcommand\p@subfigure{\thefigure\,}

\makeatother

\usepackage{fancyhdr}

\fancypagestyle{empty}{\fancyfoot[C]{\vspace*{-1.8\baselineskip}\thepage}}
\fancypagestyle{plain}{\fancyfoot[C]{\vspace*{-1.8\baselineskip}\thepage}}

\usepackage[first=0,last=9]{lcg}
\usepackage{colortbl}
\definecolor{Gray}{gray}{0.8}

\usepackage{url}

\usepackage{hyperref}
\usepackage{breakurl}

\usepackage{amsmath,amsfonts,bm}

\def\vv{{\bm{v}}}

\def\vf{{\bm{f}}}

\def\vv{{\bm{v}}}

\def\eqref#1{equation~\ref{#1}}

\def\1{\bm{1}}

\def\vf{{\bm{f}}}

\def\vv{{\bm{v}}}

\DeclareMathAlphabet{\mathsfit}{\encodingdefault}{\sfdefault}{m}{sl}
\SetMathAlphabet{\mathsfit}{bold}{\encodingdefault}{\sfdefault}{bx}{n}

\def\gL{{\mathcal{L}}}

\DeclareMathOperator{\sign}{sign}

\def\tablename{Table}
\def\figurename{Figure}

\makeatletter
\newif\if@restonecol
\makeatother

\usepackage[boxed, ruled, vlined, linesnumbered]{algorithm2e}
\SetKwRepeat{Do}{do}{while}

\newenvironment{changemargin}[2]{\begin{list}{}{
	\setlength{\topsep}{0pt}\setlength{\leftmargin}{0pt}
	\setlength{\rightmargin}{0pt}
	\setlength{\listparindent}{\parindent}
	\setlength{\itemindent}{\parindent}
	\setlength{\parsep}{0pt plus 1pt}
	\addtolength{\leftmargin}{#1}\addtolength{\rightmargin}{#2}
	}\item}
	{\end{list}}

\newcommand{\msec}[1]{\S\ref{#1}}
\newcommand{\meq}[1]{Eq.\,(\ref{#1})}
\newcommand{\mcite}[1]{~\cite{#1}}
\newcommand{\mref}[1]{\,\ref{#1}}

\usepackage{xr}

\makeatletter
\newcommand*{\addFileDependency}[1]{
  \typeout{(#1)}
  \@addtofilelist{#1}
  \IfFileExists{#1}{}{\typeout{No file #1.}}
}
\makeatother

\usepackage{scalerel}[2016/12/29]

\usepackage{diagbox}

\usepackage{array}
\newcolumntype{P}[1]{>{\centering\arraybackslash}p{#1}}

\makeatletter
\providecommand{\leadsfrom}{%
  \mathrel{\mathpalette\reflect@squig\relax}%
}
\newcommand{\reflect@squig}[2]{%
  \reflectbox{$\m@th#1\leadsto$}%
}
\makeatother

\makeatletter                 
\g@addto@macro\maketitle{\thispagestyle{plain}}
\makeatother

\usepackage{footnote}
\makesavenoteenv{tabular}
\usepackage{tabularx}

\newcommand{\norm}[1]{\left\lVert#1\right\rVert}
\DeclareMathAlphabet\mathbfcal{OMS}{cmsy}{b}{n}

\newcommand{\system}{{MSF-ADV}\xspace}
\newcommand{\DemoWeb}{\textbf{\url{https://sites.google.com/view/cav-sec/msf-adv}}}

\iftrue

\newcommand{\nsection}[1]{\section{#1}}
\newcommand{\nsubsection}[1]{\subsection{#1}}

\else

\newcommand{\nsection}[1]{\vspace{-0.1cm}\section{#1}\vspace{-0.1cm}}
\newcommand{\nsubsection}[1]{\vspace{-0.15cm}\subsection{#1}\vspace{-0.1cm}}

\fi

\usepackage{mdframed}
\mdfdefinestyle{rebuttalstyle}{innerleftmargin=3pt, innerrightmargin=3pt, innertopmargin=3pt, innerbottommargin=3pt}

\newcounter{response}[section]

\newcounter{revision}[section]

\newcounter{link}[section]

\newcounter{comments}[section]

\begin{document}

\title{Invisible for both Camera and LiDAR:  Security of Multi-Sensor Fusion based Perception in Autonomous Driving Under Physical-World Attacks}

\vspace{-1.0cm}

\author{{\rm Yulong Cao}$^{\star, \mathsection}$ \thanks{$^{\star}$Alphabetical ordering; The first four authors contributed equally.} \quad {\rm Ningfei Wang}$^{\star, \dagger}$ \quad {\rm Chaowei Xiao}$^{\star, \|, \ddagger\ddagger }$ \quad {\rm Dawei Yang}$^{\star, \mathsection}$ \quad {\rm Jin Fang}$^\ddagger$ \quad {\rm Ruigang Yang}$^{\dagger\dagger}$ 
\\
{\rm Qi Alfred Chen}$^\dagger$ \quad {\rm Mingyan Liu}$^{\mathsection}$ \quad {\rm Bo Li}$^\mathparagraph$ \\
$^\dagger$University of California, Irvine, \{ningfei.wang, alfchen\}@uci.edu\\

$^\mathsection$University of Michigan, \{yulongc, ydawei, mingyan\}@umich.edu\\
$^\|$NVIDIA Research \ \ \ \ \ $^{\ddagger\ddagger}$Arizona State University  \ \ \ \ \ $^{\dagger\dagger}$Inceptio \\
$^\ddagger$Baidu Research and National Engineering Laboratory of Deep Learning Technology and Application, China \\
$^\mathparagraph$University of Illinois at Urbana-Champaign, lbo@illinois.edu

}

\date{}

 \maketitle
\thispagestyle{plain}
\pagestyle{plain}

\thispagestyle{empty}

\bstctlcite{IEEEexample:BSTcontrol}

\begin{abstract}
In Autonomous Driving (AD) systems, perception is both security and safety critical. Despite various prior studies on its security issues, \textit{all} of them only consider attacks on camera- or LiDAR-based AD perception \textit{alone}. However, production AD systems today predominantly adopt a Multi-Sensor Fusion (MSF) based design, which in principle can be more robust against these attacks under the assumption that not all fusion sources are (or can be) attacked at the same time. In this paper, we present the first study of security issues of MSF-based perception in AD systems. We directly challenge the basic MSF design assumption above by exploring the possibility of attacking \textit{all} fusion sources simultaneously. This allows us for the first time to understand how much security guarantee MSF can fundamentally provide as a general defense strategy for AD perception.

We formulate the attack as an optimization problem to generate a physically-realizable, adversarial 3D-printed object that misleads an AD system to fail in detecting it and thus crash into it. To systematically generate such a physical-world attack, we propose a novel attack pipeline that addresses two main design challenges: (1) non-differentiable target camera and LiDAR sensing systems, and (2) non-differentiable cell-level aggregated features popularly used in LiDAR-based AD perception.
We evaluate our attack on MSF algorithms included in representative open-source industry-grade AD systems in real-world driving scenarios. Our results show that the attack achieves over 90\% success rate across different object types and MSF algorithms. Our attack is also found stealthy, robust to victim positions, transferable across MSF algorithms, and physical-world realizable after being 3D-printed and captured by LiDAR and camera devices.
To concretely assess the end-to-end safety impact, we further perform simulation evaluation and show that it can cause a 100\% vehicle collision rate for an industry-grade AD system. We also evaluate and discuss defense strategies.

\end{abstract}

\section{Introduction}
\label{sec:introduction}

Today, high-level (e.g., Level-4~\cite{sae2018}) self-driving cars, or Autonomous Vehicles (AV)~\cite{av-companies}, are under rapid development. Some of them, e.g., Google Waymo~\cite{waymo-one} and TuSimple~\cite{tusimple-truck}, are already providing services on public roads. To ensure correct and safe driving, a fundamental pillar in the Autonomous Driving (AD) system is \textit{perception}, which leverages sensors such as cameras and LiDARs (Light Detection and Ranging) to detect surrounding obstacles in real time. Due to the direct impact on safety-critical driving decisions such as collision avoidance, various prior works have studied the security of AD perception under realistic physical-world attack vectors such as adding stickers, posters, or paintings to traffic signs\mcite{eykholt2018physical, zhao2018seeing, lu2017adversarial, zhangcamou, chen2018shapeshifter}, or shooting lasers to the LiDAR\mcite{cao2019adversarial, jiachen:usenix:2020}.

All of these studies, however, are limited to attacks on a single source of AD perception, i.e., camera- or LiDAR-based AD perception \textit{alone}\mcite{eykholt2018physical, zhao2018seeing, lu2017adversarial, zhangcamou, chen2018shapeshifter, pei2017deepxplore, tian2018deeptest, cao2019adversarial, jiachen:usenix:2020}. By contrast, production high-level AD systems such as Waymo, Pony.ai, and Baidu Apollo, typically adopt a Multi-Sensor Fusion (MSF) based design \mcite{waymo-msf,apollo, autoware, ponyai-msf}, which fuses the results from different perception sources, e.g., LiDAR and camera, to achieve overall higher accuracy and robustness\mcite{frossard2018end,liang2018deep, chen2017multi,xu2018pointfusion,liang2019multi,du2017car, ku2018joint,ma2019accurate, du2018general
}. In such a design, under the assumption that not all perception sources are (or can be) attacked simultaneously, there \textit{always} exists a possible MSF algorithm that can rely on the unattacked source(s) to detect or prevent such an attack. This basic security design assumption is believed to hold in general \mcite{quinonezsavior, guo2018roboads}, and MSF is thus widely recognized as a general defense strategy against existing attacks on AD perception\mcite{quinonezsavior, guo2018roboads, cao2019adversarial, shin2017illusion}.

This paper presents a first study on the security property of MSF-based perception in AD systems today. We directly challenge the above basic security design assumption by demonstrating the possibility of effectively and simultaneously attacking \textit{all} perception sources used in state-of-the-art MSF-based AD perception, i.e., camera and LiDAR\mcite{frossard2018end,liang2018deep, chen2017multi,xu2018pointfusion,liang2019multi,du2017car, ku2018joint,ma2019accurate, du2018general
}. This for the first time allows us to gain a concrete understanding of how much security guarantee the use of MSF can fundamentally provide as a general defense strategy for AD perception. Specifically, we consider physical-world attack vectors for high attack practicality, and target an attack goal with the most direct safety consequence for autonomous driving: cause a victim AV to fail in detecting a front obstacle.

Although prior works have designed successful physical-world attacks on AD perceptions based only on camera or only on LiDAR, we find that simply combining their designs does not achieve our goal. First, we need to identify a physical-world attack vector effective for both camera and LiDAR, which can not be satisfied by those popular ones used in prior works. For example, adding stickers changes an object's texture (e.g., color) but not its shape; this can be effective for camera\mcite{eykholt2018physical, zhao2018seeing, lu2017adversarial, zhangcamou, chen2018shapeshifter} but not LiDAR. Conversely, laser shooting has been shown to be effective for LiDAR-based AD perception\mcite{cao2019adversarial, jiachen:usenix:2020}, but not for camera-based ones. Second, no matter what attack vector we use, we need to further address 2 design challenges: (1) We need to differentiably synthesize the physical attack impacts simultaneously and consistently onto both camera images and LiDAR point clouds.
For certain attack vectors, e.g., differentiably modelling the impact of lasers on camera images, this can be very challenging. (2) To improve  run-time performance, the state-of-the-art LiDAR-based AD perception uses aggregated features of the 3D points grouped at the level of 2D or 3D cells\mcite{engelcke2017vote3deep,zhou2018voxelnet,lang2019pointpillars, apollo,beltran2018birdnet,yang2018pixor,chen2017multi}; however, their calculation is by nature non-differentiable (\S\ref{sec:design-challenge}), which makes the attack difficult to optimize.

Towards this end, we design a novel physical-world adversarial attack method, \system, which addresses the challenges above and thus fundamentally challenges the basic MSF design assumption in AD perception. We employ \textit{adversarial 3D object} as the attack vector, with the key observation that different shapes of a 3D object can lead to both point position changes in LiDAR point clouds and pixel value changes in camera images. Thus, an attacker can leverage \textit{shape manipulations} to introduce input perturbations to both camera and LiDAR simultaneously. To achieve the attack goal, the attacker simply places such an object on the roadway; this can be conveniently accomplished with modern 3D printing services and an object type commonly expected on the roadway, e.g., a traffic cone but with a slightly worn or broken look.

To systematically generate effective adversarial 3D objects, we adopt an optimization-based approach that starts with a 3D mesh of a normal object, e.g., a normal traffic cone, and performs shape manipulation by changing its vertex positions. Under this attack vector, we address design challenge 1 by constructing differentiable 3D rendering functions to synthesize the attack-influenced point clouds and camera images. For design challenge 2, we find that all commonly-used cell-level aggregated features can be differentiably derived by the \textit{point-inclusion} property (\S\ref{sec:preprocessing}). Thus, we first design a differentiable and accurate approximation for such property, and then use it as a building block to differentiably compute the gradient of the cell-level aggregated features during the optimization. We also employ domain-specific designs for attack robustness, stealthiness, and physical-world realizability.

We evaluate \system with MSF algorithms included in 2 open-source full-stack AD systems, Baidu Apollo\mcite{apollo} and Autoware.AI\mcite{autoware}, that have high representativeness in practice, e.g., Apollo is ranked as the top 4 leading AD developers along with Waymo, Ford, and Cruise\mcite{ranking-AD}. We select 3 object types and evaluate each on 100 real-world driving scenarios from the KITTI dataset\mcite{kittidataset}. Our results show that the generated adversarial objects achieve more than 91\% success rate across different object types and MSF algorithms. We also find that our attack is (1) stealthy from the driver's view based on a user study, (2) robust to different victim approaching positions and angles, with over 95\% average success rates, and (3) transferable across different MSF algorithms, with an average transfer attack success rate of around 75\%.

To understand the attack realizability in the physical world, we 3D-print our adversarial objects, and evaluate them using real LiDAR and camera devices. Using a vehicle with a LiDAR mounted, we find that our 3D-printed adversarial object can successfully evade LiDAR detection in 99.1\% (107) of the total 108 collected frames. Using a miniature-scale experiment setting (\S\ref{sec:miniature-scale-attack-eval}), we find that our adversarial object has a 85-90\% success rate to evade both LiDAR and camera detection at 20 randomly-sampled positions.

To understand the end-to-end safety impact, we further evaluate our method using a production-grade AD simulator, and find that our adversarial traffic cone can cause a 100\% vehicle collision rate for an Apollo AV across 100 runs. In contrast, the collision rate with a normal traffic cone is 0\%. Demo videos are at our project website: \DemoWeb. We also evaluate various existing DNN-level defense strategies (e.g., input transformation and augmenting training data), and discuss future defense directions. Our code and data are released at our website~\cite{ourwebsite}.

In summary, this work makes the following contributions:
\begin{itemize}
    \item We are the first to study security issues of MSF-based AD perception and the first to challenge the basic MSF design assumption in the AD context. We successfully design and engineer a physical-world adversarial attack aiming at generating adversarial 3D object to mislead a victim AV to fail in detecting it and thus crash into it.
    \item We adopt an optimization-based approach that addresses two main design challenges: non-differentiable target camera and LiDAR sensing systems, and non-differentiable cell-level aggregated features used by LiDAR. We also design strategies to enhance the attack robustness, stealthiness, and physical-world realizability.
    \item We evaluate on MSF algorithms included in representative open-source industry-grade AD systems in real-world driving scenarios. Our attack is shown to achieve over 91\% success rates across different object types and MSF algorithms. Such high effectiveness can also be achieved with (1) high stealthiness, (2) high robustness to victim positions, (3) high transferability across MSF algorithms, and (4) high physical-world realizability after being 3D-printed and captured by LiDAR and camera devices.
    \item To understand the end-to-end safety impact, we further evaluate the proposed attack on a production-grade simulator, and show that our attack can cause a 100\% vehicle collision rate to an industry-grade AD system. We also evaluate and discuss defense strategies.
\end{itemize}

While MSF is widely recognized as a promising and general defense strategy for existing attacks on AD perception\mcite{quinonezsavior, guo2018roboads, cao2019adversarial, ivanov2014attack, shin2017illusion, xu2018analyzing, yan2016can, petit2014potential, man2020ghostimage, rearchitecting}, prior works have neither studied the security of existing MSF algorithms in practical AD settings, nor made attempts to understand whether the very basic security design assumption for MSF can fail. In this paper, we make the first attempt towards this direction, and we hope that our findings and insights can inspire more future research into this largely overlooked research perspective.

\section{Background}
\label{sec:background}

\subsection{MSF-based AD Perception}
\label{sec:background-msf}
In high-level (e.g., Level 4\mcite{sae2018}) AD systems, \textit{perception} is a critical module that detects surrounding objects in real time.
Due to its direct impact on safety-critical driving decisions such as collision avoidance, AD perception in production high-level AD systems such as Google Waymo, Pony.ai, and Baidu Apollo predominantly adopts a Multi-Sensor Fusion (MSF) based design\mcite{waymo-msf,apollo, autoware, ponyai-msf}. In this paper, we call such design \textit{MSF-based AD perception}, or \textit{MSF} for short. In this paper, we focus on MSF designed for \textit{in-road obstacle detection}, e.g., front cars, which is the most basic task for AD perception.

\textbf{MSF design principle and basic assumption.} In MSF-based AD perception, the final object detection results are obtained by fusing multiple perception sources such as camera and LiDAR, with the goal of leveraging their strengths while compensating their weaknesses to achieve overall higher accuracy and robustness than those achievable by a single perception source\mcite{frossard2018end,liang2018deep, chen2017multi,xu2018pointfusion,liang2019multi,du2017car, ku2018joint,ma2019accurate, du2018general
}. For example, LiDAR is a ranging-based sensor by shooting lasers, which thus is more difficult to capture the texture information (e.g., color) of an object compared to cameras\mcite{frossard2018end}. Camera images, on the other hand, cannot directly provide the depth information of an object\mcite{liang2018deep,ma2019accurate}, which can be overcome by LiDARs. Thus, an MSF algorithm can be designed to leverage both the depth information from LiDAR point clouds and the texture information from camera images to achieve higher object detection performance than those using either camera or LiDAR alone\mcite{chen2017multi,liang2018deep}. To achieve such overall higher accuracy and robustness, the basic design assumption is that \textit{there generally exists at least one source that can provide the correct results}. In this paper, we are the first to challenge such assumption in the AD context.

\begin{figure}[t]
    \footnotesize
      \centering
          \includegraphics[width=0.98\linewidth]{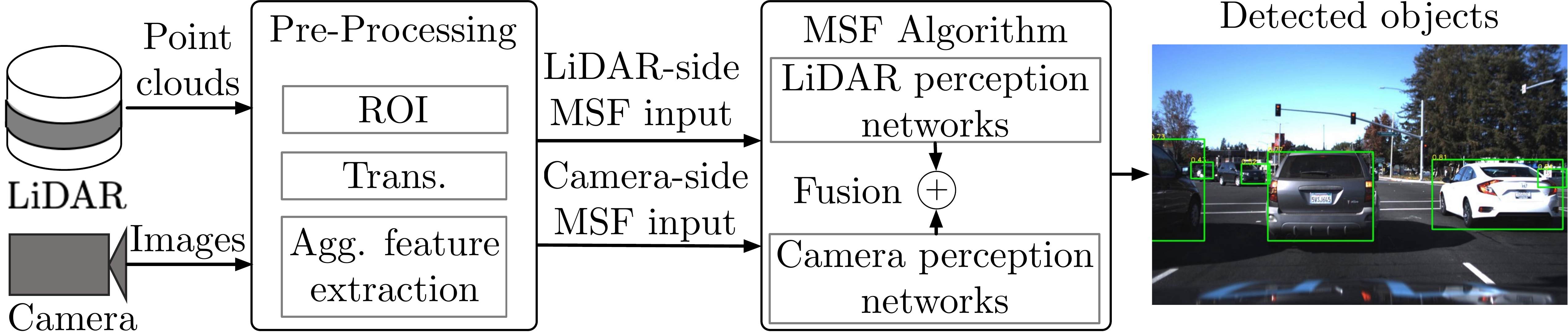}  
    \caption{Overview of MSF-based AD perception design.}
        \label{fig:framework-msf}
\end{figure}

{\bf Representative MSF algorithm design.} In AD perception, state-of-the-art MSF algorithms predominately use 2 perception sources: camera and LiDAR\mcite{frossard2018end,liang2018deep, chen2017multi,xu2018pointfusion,liang2019multi,du2017car, ku2018joint, du2018general
}. Fig.\mref{fig:framework-msf} shows an overview of a typical MSF-based AD perception design. In industry AD systems, before running the MSF, the raw camera and LiDAR inputs, i.e., camera images and LiDAR point clouds, are usually first pre-processed\mcite{apollo, autoware} to prepare the camera- and LiDAR-side MSF inputs, which can improve the run-time algorithm performance (detailed later).

In the MSF algorithm, state-of-the-art designs predominantly adopt DNN networks to process the LiDAR-side and camera-side MSF inputs\mcite{apollo, autoware, frossard2018end,liang2018deep, chen2017multi,xu2018pointfusion,liang2019multi,du2017car, ku2018joint, du2018general}, due to the recent superior performance of deep learning in object detection\mcite{obj-detect-perf}. In this paper, we call them \textit{LiDAR perception networks} and \textit{camera perception networks} inside the MSF algorithm. Next, the processing results from these two networks are fused using (1) DNNs\mcite{chen2017multi, xu2018pointfusion, frossard2018end, liang2018deep, liang2019multi, du2017car, ku2018joint}, or (2) hard-coded matching and prioritization rules\mcite{apollo, autoware}. Rule-based fusion is usually a late fusion, i.e., applied to the end results of the two networks, while DNN-based one can be a late or early fusion, i.e., at the intermediate perception results, which can be fused more deeply and thus potentially lead to higher accuracy. Meanwhile, rule-based fusion has two unique benefits. First, it is more modular and thus can flexibility combine different camera and LiDAR perception models\mcite{modular-end2end}. Second, it is easier to debug and interpret than DNNs\mcite{chi1708deep}, and also to hard-code safety rules and measures\mcite{modular-end2end}. In our attack design later in \S\ref{sec:methods}, we comprehensively consider both fusion designs.

When preparing the camera- and LiDAR-side MSF inputs, typical pre-processing steps include data transformation such as rotations and shifting, applying Region of Interest (ROI) filter to remove unrelated input portions, and extracting aggregated features from the raw input. These pre-processing steps can largely reduce the sizes and dimensions of the MSF algorithm inputs, which can thus greatly improve the run-time algorithm performance\mcite{pcsurvey}. Considering that the raw point cloud data can include millions of 3D points per second\mcite{lidar-128}, such pre-processing is especially beneficial for LiDAR perception. Thus, many state-of-the-art LiDAR-based AD perception model designs choose to use aggregated input features such as average height and intensity of the 3D points grouped at the level of \textit{3D cells}, or \textit{voxels}\mcite{engelcke2017vote3deep,wang2015voting,zhou2018voxelnet,lang2019pointpillars}. 
Some state-of-the-art designs even choose to further aggregate the features in such 3D cells to 2D cells in Bird's-Eye View (BEV) to further improve the real-time detection performance\mcite{yang2018pixor}, which is thus the most popularly adopted in industry-grade AD systems\mcite{apollo,beltran2018birdnet,yang2018pixor,chen2017multi}. As detailed in~\S\ref{sec:design-challenge}, such popular adoption of cell-level aggregated features for LiDAR introduces a unique challenge to our attack design.

\subsection{Physical-World Adversarial Attack}
\label{sec:adv-attack}

Recent works find that DNN models are generally vulnerable to adversarial example, or \textit{adversarial attacks}\mcite{goodfellow:fsgm, papernot:jsma,xiao2018generating,xiao2018spatially,xiao2018characterizing,xiao2019characterizing,qiu2020semanticadv}. Some works further explored such attacks in the physical world\mcite{eykholt2018physical, li2019adversarial,adv-patch,thys2019fooling,zhangcamou,chen2018shapeshifter,lu2017adversarial,zhao2018seeing,athalye2017synthesizing}. In the AD context, previous works have designed successful physical-world adversarial attacks on the camera-based AD perception \textit{alone}\mcite{eykholt2018physical, zhao2018seeing, lu2017adversarial, zhangcamou, chen2018shapeshifter, pei2017deepxplore, tian2018deeptest}, or the LiDAR-based one \textit{alone}\mcite{cao2019adversarial, jiachen:usenix:2020}. However, none of them have considered MSF-based AD perception, which is predominantly adopted in industry AD systems today (\S\ref{sec:background-msf}) and in principle can be more robust against these attacks (\S\ref{sec:introduction}). Also, as detailed later in \msec{sec:design-challenge}, blindly combining these prior designs cannot directly lead to successful attacks on MSF due to various new and unique challenges.

\section{Problem Formulation and Design Challenges}
\label{sec:problem_challenges}

\subsection{Attack Goal and Threat Model}
\label{sec:attack-goal}

\textbf{Attack goal: Fundamentally defeat MSF design assumption.} In this paper, we target an attack goal with the most direct safety impact on driving: fool the MSF-based AD perception in the victim AV to fail in detecting a front obstacle and thus crash into it. Even when the vehicle has a fail-safe Automatic Emergency Brake (AEB) system, e.g., based on RADAR or ultrasonic sensors, such a crash is still possible for two reasons. First, today's AEB systems are not perfect. For example, a recent study shows that the ones in popular vehicle models today fail to avoid crashes 60\% of the time~\cite{aeb-fail}. Second, even if they can successfully perform emergency stop, they cannot avoid being hit by rear vehicles that fail to yield on time. To achieve this goal, in this paper we target physical-world attack vectors in the AD context for high practicality and realism. 

Due to the basic design assumption of MSF (\S\ref{sec:background-msf}), as long as there still exists at least one perception source that is not attacked, it is always possible for the unattacked source(s) to correct the final fused perception results and thus defeat our attack goal. Thus, in this paper we aim at designing an attack that can effectively attack \textit{all} perception sources used in the MSF-based AD perception. This can enable our design to \textit{fundamentally} defeat the MSF design assumption and thus most generally achieve our goal above. As the combination of camera and LiDAR is most popularly adopted in state-of-the-art MSF-based AD perception (\S\ref{sec:background-msf}), in this paper our design needs to attack both camera and LiDAR simultaneously.

\label{sec:threat-model}

\textbf{Threat Model.} As the first study to achieve the attack goal above, in this work we mainly focus on a white-box attack setting, i.e., assuming that the attacker has a full knowledge of the MSF algorithm used in the victim AD system. This is the same assumption made in most prior adversarial attacks on camera- or LiDAR-based AD perception\mcite{cao2019adversarial, eykholt2018robust, eykholt2018physical, zhao2018seeing}.
To achieve this, the attacker may obtain a victim AV model, e.g., by purchasing or renting\mcite{avis-waymo}, and then reverse engineer its perception module, which has been shown as possible on Tesla Autopilot\mcite{reverse-tesla}.
The attacker can also target the AVs using open-source MSF-based AD perception algorithms\mcite{apollo, autoware}. In the attack preparation time, we assume that the attacker can collect camera images and LiDAR point clouds of a targeted road where she plans to launch the attack.

\vspace{-0.15cm}
\subsection{Design Challenges}
\label{sec:design-challenge}

As described in~\S\ref{sec:background-msf}, in state-of-the-art MSF algorithms, the camera and LiDAR perception networks are DNN based. Although no prior works consider attacking MSF, many designed successful physical-world adversarial attacks on camera- \textit{or} LiDAR-based AD perception DNN models. However, we find that blindly combining these prior designs cannot directly achieve our goal due to 3 unique challenges:

{\bf C1. Lack of a single physical-world attack vector effective for both camera- and LiDAR-based AD perception.} To achieve our attack goal, we need to find physical-world attack vectors for both camera- and LiDAR-based perception networks in MSF. However, so far none of the attack vectors used in previous physical-world adversarial attacks in the AD context have shown effectiveness in affecting both. For camera-based AD perception, previous works predominately consider adding stickers/posters~\cite{eykholt2018physical, zhao2018seeing}, painting~\cite{zhangcamou, chen2018shapeshifter}, or changing brightness~\cite{pei2017deepxplore, tian2018deeptest}, which can only change the \textit{texture} of an obstacle but not its \textit{shape} and thus can barely affect the LiDAR point clouds. On the LiDAR side, LiDAR spoofing~\cite{cao2019adversarial, jiachen:usenix:2020}, which shoots lasers to LiDAR, has shown to be effective in the AD context. Although lasers can also affect camera inputs~\cite{yan2016can}, no prior work has studied its effectiveness for fooling camera-based AD perception models. One possible solution is to use separate attack vectors for them, e.g., using stickers for camera and laser shooting for LiDAR. However, this not only adds up the attack deployment costs and thus lowers the realizability and stealthiness, but also requires precise synchronizations across the attack processes. Thus, it is highly desired to identify one single attack vector that can effectively attack both at the same time.

{\bf C2. Need to differentiably synthesize physically-consistent attack impacts onto both camera and LiDAR.} To systematically generate adversarial inputs, prior works generally adopt optimization-based approaches, which have shown both high efficiency and effectiveness\mcite{zhao2018seeing,athalye2017synthesizing}. Since adversarial attack generation typically takes thousands of optimization iterations\mcite{madry:towards, carlini:cw}, it is almost impossible in practice to physically drive vehicles on the target road to obtain the attack-influenced camera images and LiDAR point clouds every time the adversarial inputs are updated in an iteration. Thus, we need to digitally synthesize the impacts of the adversarial stimulus from the physical world onto both camera images and LiDAR point clouds, and such synthesizing needs to be differentiable to enable effective optimization. As discussed in \textbf{C1}, no single attack vector has been studied for both camera- and LiDAR-based AD perception so far in prior works. Thus, no matter what attack vector we identify to address \textbf{C1}, we need to design a new differentiable synthesizing function for at least one of the perception sources, which can be quite challenging for certain physical-world attack vectors, e.g., differentiably modelling the impact of lasers on camera inputs from different distances and angles. Meanwhile, since such attack impacts come from the same physical-world stimulus, the synthesized impacts to the camera images and the LiDAR point clouds need to be \textit{physically consistent}, e.g., conforming to their different mounting positions in the AV.

{\bf C3. Need to handle non-differentiable pre-processing steps in AD perception.} As introduced in~\S\ref{sec:background-msf}, in industry AD systems, images and point clouds are usually pre-processed before fed into the MSF algorithm. In particular, state-of-the-art LiDAR-based AD perception models popularly use aggregated features of 3D points grouped at level of 2D or 3D cells (\S\ref{sec:background-msf}). To calculate such cell-level aggregated features, the necessary first step is to calculate whether an input point is inside a cell or not. In this paper, we call it a \textit{point-inclusion} property. By nature, such property is discontinuous, i.e., 0 and 1 for outside and inside a cell. This causes the calculation of any cell-level aggregated features non-differentiable with regard to the LiDAR point clouds, which thus makes our optimization difficult to be effective. So far, no prior works have considered a general design to handle such non-differentiable pre-processing steps for LiDAR.

\section{Attack Design: MSF-ADV}
\label{sec:methods}

In this paper, we are the first to address all the 3 challenges in~\S\ref{sec:design-challenge} by designing a novel physical-world adversarial attack method, \textit{\system}, which thus can fundamentally defeat the MSF design assumption in AD perception.

\subsection{Design Overview}
\label{sec:design-overview}
\vspace{0.02cm}

To address the challenges in~\S\ref{sec:design-challenge}, our \system method has the following novel designs:

{\bf Adversarial 3D object: physically-realizable and stealthy attack vector for MSF-based AD perception.}
To address \textbf{C1}, we identify \textit{adversarial 3D object} as the physical-world attack vector against MSF-based AD perception. Our key insight is that different shapes of a 3D object can lead to not only point position changes in LiDAR point clouds but also pixel value changes in camera images. Thus, the attacker can leverage \textit{shape manipulations} of such an object to introduce adversarial input perturbations simultaneously to both camera and LiDAR perception networks in the MSF algorithm. To achieve the attack goal, the attacker simply places such an object in the roadway to trick the victim AV to crash into it. 

\begin{figure}[t]
      \centering
          \includegraphics[width=0.8\linewidth, height=2.5cm]{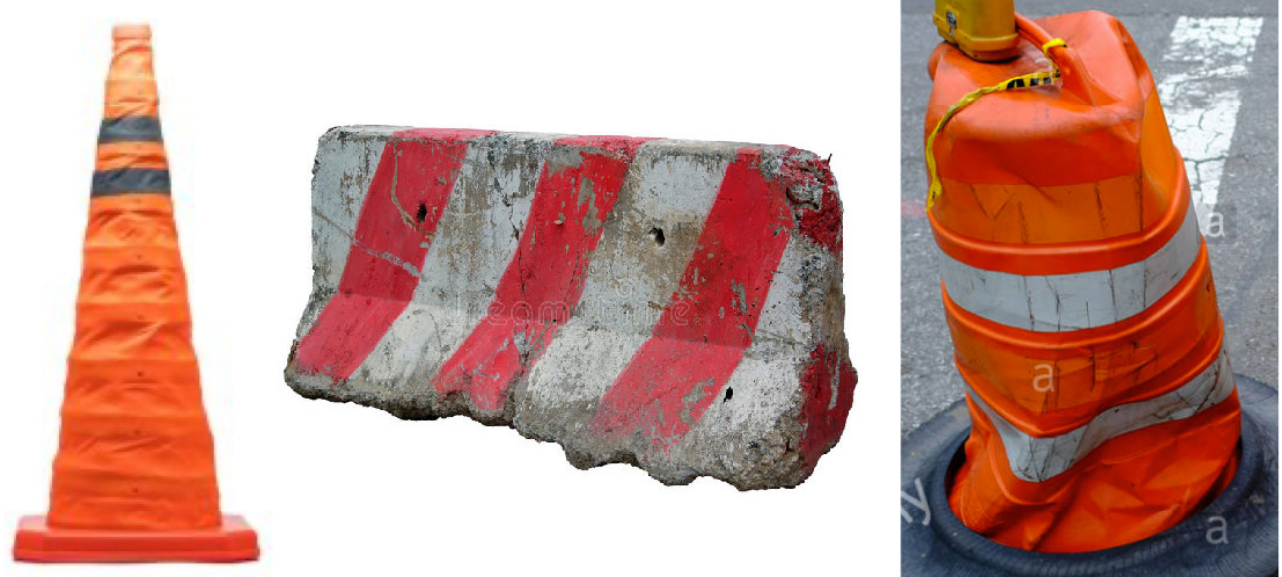}
    \caption{Real-world traffic objects with worn or broken looking shapes, which can be mimicked by our physical-world attack vector: adversarial 3D object with \textit{shape manipulations}.}
        \label{fig:attack-vectors}
\end{figure}

Beside satisfying \textbf{C1}, such an attack vector also has 2 other advantages. First, it is easily realizable and deployable in the physical world. For example, the attacker can construct it digitally in a 3D mesh and 3D-print it, which is convenient today through online services\mcite{3d-print}. Second, it can achieve high stealthiness by mimicking a normal traffic object that can legitimately appear in the middle of the road, e.g., a traffic cone or barrier, but with a worn or broken look, which is not uncommon in the real world as shown in Fig.\mref{fig:attack-vectors}. In our design (\S\ref{sec:objective-function}), we also constrain the degree of the shape changes from the normal object to achieve high stealthiness. Note that although it is possible to manipulate the object texture (e.g., color) together with the shape in our design, we intentionally choose to not consider it in this paper as it can greatly harm stealthiness and also incur additional printability issues, which is a common challenge for physical-world adversarial attacks using stickers/posters~\cite{eykholt2019designing, eykholt2018physical, huang2020spaa}.

\textbf{Causing road safety threats.} To make such an object both easy to deploy and able to cause severe crashes, the attacker can choose smaller objects such as a rock or traffic cone but fill it with granite or even metal to make it harder and heavier. For example, a 0.5 cubic-meter rock or a 1-meter high traffic cone~\cite{traffic-cone-1m} filled with some aluminum can easily weigh over 100 kg, which can trip the victim AV to lose control, damage the chassis, or break the windshield glass if bounced up when driving at a high speed. Besides causing damages by the crash itself, the attackers can also exploit the \textit{semantic meaning} of certain road object types such as traffic cones. For example, the attacker can design an AV-specific attack by placing nails or glass debris behind an adversarial traffic cone object so that failing to detect it can lead to tire blowout of a targeted AV. Here, the safety damages are not directly caused by the traffic cone crash itself, and thus in this case the adversarial traffic cone can be small and lightweight like normal ones to make it easier to 3D-print, carry, and deploy.

\begin{figure*}[t]
      \centering
          \includegraphics[width=.96\linewidth]{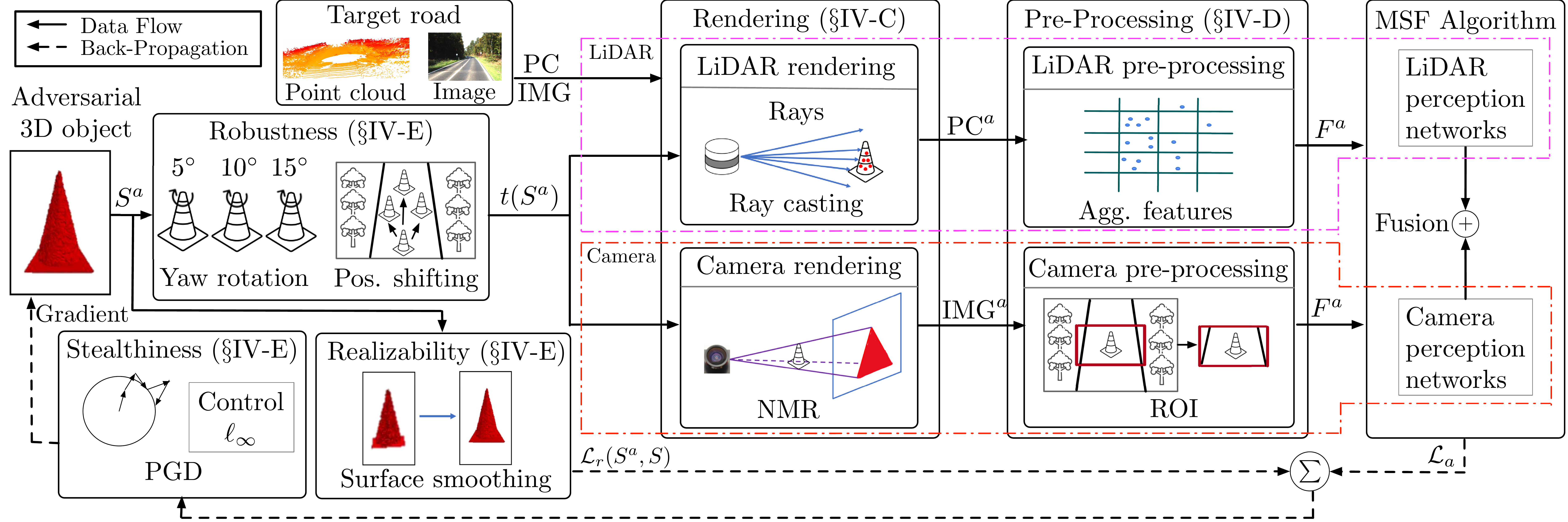}  
    
	\caption{Overview of the optimization-based adversarial 3D object generation in \system.}
		\label{fig:framework}
\end{figure*}

{\bf Optimization-based adversarial 3D object generation.} To systematically generate adversarial 3D objects, we adopt an optimization-based approach similar to prior works\mcite{eykholt2018physical, zhao2018seeing, lu2017adversarial, zhangcamou, chen2018shapeshifter, cao2019adversarial, jiachen:usenix:2020}. We start with a 3D mesh of a normal 3D object, e.g., a normal traffic cone, and then introduce shape manipulations by changing its vertex positions. To address \textbf{C2}, due to the choice of adversarial 3D objects as the attack vector, we can conveniently leverage existing 3D rendering techniques in computer graphics to simulate the functionalities of the physical equipment, i.e., camera and LiDAR, and thus systematically synthesize the attack-influenced camera images and LiDAR point clouds. Specifically, to enable the end-to-end optimization process, we perform differentiable constructions of these rendering processes, and use the relative positions to the 3D object to ensure the physical consistency with the corresponding camera and LiDAR mounting positions.

With the synthesized raw camera images and LiDAR point could, next we design the differentiable  approximation function for the non-differentiable pre-processing step (non-differentiable cell-level aggregated feature calculation) to enable the end-to-end optimization. To address this, our key insight is that all the commonly-used cell-level aggregated features can be differentiably derived by the point-inclusion property (detailed later in~\S\ref{sec:preprocessing}). Thus, we first design a novel and accurate differentiable function to approximate the calculation of the point-inclusion property, and then use it as a building block to achieve differentiable computations of the pre-processing steps for LiDAR. In the optimization process, we also have other domain-specific designs, e.g., for attack robustness, stealthiness, and physical-world realizability, which will be detailed in the following sections.

\subsection{\system Methodology Overview}
\label{sec:opt-proc-overview}

In this section, we provide an overview of our \system method, and will detail its components in later sections. 

\textbf{Problem formulation.} We formulate the attack generation process as the following optimization problem:

\small
\begin{align}
    \min_{S^a}\ \ & \mathbb{E}_{t\sim T}[\gL_a(t(S^{a} ); \mathcal{R}^{l}, \mathcal{R}^{c}, \mathcal{P}, \mathcal{M})  + \lambda \cdot \gL_{r}(S^a, \ S)]
    \label{eq:overview-opt}\\
    {\rm \text{where}} \ \
    & \mathrm{PC}^a = \mathcal{R}^l( t( S^a), \ \mathrm{PC})  \label{eq:overview-lidar}\\
    & \mathrm{IMG}^a = \mathcal{R}^c( t(\ S^a), \ \mathrm{IMG}, \mathrm{C}) \label{eq:overview-camera}\\
    & F^a = \mathcal{P}(\mathrm{PC}^a, \ \mathrm{IMG}^a) \label{eq:overview-preprocessing}\\
    & \gL_a(t(S^{a} ); \mathcal{R}^{l}, \mathcal{R}^{c}, \mathcal{P}, \mathcal{M}) = \mathcal{O}(\mathcal{M}(F^a)) \label{eq:overview-msf-alg}\\
     {\rm \text{subject to}} \ \ 
    &  \Delta(S^a, \ S)  \leq \epsilon \label{eq:overview-pgd}
\end{align}
\normalsize

$S$ is the original benign object and $S^{a}$ is the adversarial one. We use vertex-face ($\vv$-$\vf$) meshes to represent them, i.e., $S=(\vv, \vf)$ and $S^a=(\vv^a, \vf^a)$. 
In \meq{eq:overview-opt}, the optimizing parameter is the adversarial object $S^a$, and we only change its vertices $\vv^a$.
The objective function includes: (1) an adversarial loss $\gL_{a}$, which is designed to achieve our attack goal by misleading the MSF algorithm $\mathcal{M}(\cdot)$ to fail in detecting $S^{a}$, and (2) a realizability loss $\gL_r(\cdot)$, which is designed to improve smoothness of the $S^{a}$ surface to benefit both the printability and stealthiness (\S\ref{sec:objective-function}). To improve the robustness of $S^{a}$ in the physical world, we apply Expectation over Transformation (EoT)\mcite{athalye2017synthesizing} by introducing a set of 3D transformation $T$ to $S^{a}$ and optimize the expectation of their objective function values in \meq{eq:overview-opt}. $\lambda$ is a balancing hyper-parameter.

In \meq{eq:overview-lidar} and \meq{eq:overview-camera}, $\mathcal{R}^l(\cdot)$ and $\mathcal{R}^c(\cdot)$ are the differentiable LiDAR and camera rendering functions respectively (\S\ref{sec:rendering}). They generate the attack-influenced point clouds $\mathrm{PC}^a$ and images $\mathrm{IMG}^a$ given the corresponding backgrounds of the target road ($\mathrm{PC}$ and $\mathrm{IMG}$). $\mathrm{PC}^a$ and $\mathrm{IMG}^a$ are then fed into the differentiable pre-processing approximation function $\mathcal{P}(\cdot)$ to obtain the attack-influenced MSF input features $F^a$ (\S\ref{sec:preprocessing}). $F^a$ is fed into MSF algorithm $\mathcal{M}(\cdot)$ in \meq{eq:overview-msf-alg}, and $\mathcal{O}(\cdot)$ is designed to extract the output features related to the object's 
confidence score of the adversarial object. To achieve high stealthiness, in \meq{eq:overview-pgd} we limit the shape deformation between $S$ and $S^a$ within a threshold $\epsilon$ by using a distance metric $\Delta(\cdot)$ (e.g., $L_p$ distance metric: $\Delta(S^a, S) = || S^a - S||_{p}$).

{\bf Optimization process overview.} Fig.\mref{fig:framework} overviews our optimization process. As shown, given a 3D object $S^a$ initialized with $S$, we first apply 3D transformations (e.g. rotation and position shifting) $T$ to generate multiple samples $t(S)$ to improve the robustness of the adversarial object against environment's variation. Next, each one of them, along with the LiDAR point clouds ($\mathrm{PC}$) and camera image ($\mathrm{IMG}$) background from the target road, are fed into the rendering functions ($\mathcal{R}^{l}(\cdot), \mathcal{R}^{c}(\cdot)$), pre-processing approximation functions ($\mathcal{P}(\cdot)$), and the MSF algorithm ($\mathcal{M}(\cdot)$)  to calculate $\gL_a$. Additionally, the realizability loss $\gL_r(S^a, S)$ is added to  $\gL_a(\cdot)$ together using \meq{eq:overview-opt} to construct our loss function. To solve it, we use Projected Gradient Descent (PGD). Specifically, we compute its gradients with respect to the vertex positions $\vv^a$ of $S^a$ and constrain the gradients with a stealthiness threshold $\epsilon$. We then update $S^a$ using these gradients.  We iteratively apply this process until $S^a$ cannot be detected by the MSF algorithm. 

\subsection{Differentiable Rendering}

\label{sec:rendering}

In this section, we detail the differentiable rendering functions $\mathcal{R}^l(\cdot)$ and $\mathcal{R}^c(\cdot)$ in \meq{eq:overview-lidar} and \meq{eq:overview-camera}. To ensure physical consistency, we define $S^a$ in the LiDAR coordinate system, which is convenient as it is by nature 3D. For camera rendering, we then use a calibration matrix $C$ to transform $S^a$ from the LiDAR coordinate system to the camera coordinate system. $C$ can be obtained by measuring the relative positions between the camera and the LiDAR of AV. To achieve differential rendering, we leverage existing differentiable ray-casting methods\mcite{ray-casting-intersection} for LiDAR and NMR\mcite{rendering} for camera.

\begin{table}
\footnotesize
\centering

			\begin{tabular}{l c }
\toprule
				Cell-level Aggregated Features & Used in\\
				\midrule
				 Occupancy & \cite{apollo, yang2018pixor, autoware, engelcke2017vote3deep, wang2015voting, maturana2015voxnet}\\
			    Count & \cite{apollo, autoware}\\
				Height (min/max/mean)  & \cite{apollo, autoware, chen2017multi, beltran2018birdnet, ku2018joint}\\
				Intensity (min/max/mean)  & \cite{apollo, autoware, yang2018pixor, chen2017multi, beltran2018birdnet, engelcke2017vote3deep, wang2015voting}\\
				Density & \cite{chen2017multi, beltran2018birdnet, ku2018joint}\\
				\bottomrule
			\end{tabular}
    \caption{ Summary of commonly-used cell-level aggregated features in state-of-the-art LiDAR-based object detection models. Our novel soft point-inclusion property calculation (\S\ref{sec:preprocessing}) can be used to differentiably derive all of them.}
    
    \label{tab:features-bev-voxel}
\end{table}

\subsection{Pre-Processing Step Approximation}
\label{sec:preprocessing}

In this section, we detail the construction of the differentiable pre-processing function $\mathcal{P}(\cdot)$. Most of the pre-processing steps such as ROI, rotation, and position shifting (\S\ref{sec:background-msf}) can be directly constructed differentiably using projective and affine transformations. However, such construction is especially challenging for the calculation of cell-level aggregated features such as cell occupancy and the mean height of the points inside a cell, due to the discontinuity of the point-inclusion property as discussed in \textbf{C3} (\S\ref{sec:design-challenge}). However, such features are commonly used in state-of-the-art LiDAR-based AD perception as summarized in Table\mref{tab:features-bev-voxel}, for achieving high run-time performance (\S\ref{sec:background-msf}). This thus makes it necessary to address this to ensure the generality of our attack method.

To address this, we find that as long as we can obtain the point-inclusion value of each 3D point to a given cell, all the commonly-used features in Table\mref{tab:features-bev-voxel} can be mathematically calculated in closed form. Thus, we first design an accurate and differentiable approximation of the point-inclusion property calculation, or a \textit{soft} point-inclusion calculation, and then use it as a \textit{building block} to differentiably derive the features.

\begin{figure*}[t]
    \centering
    \begin{minipage}[t]{0.68\textwidth}
    \centering
    \vspace{-2.9cm}
     \begin{subfigure}{0.26\linewidth}
         \centering
         \vspace{-0.09cm}
         \includegraphics[width=1.\linewidth]{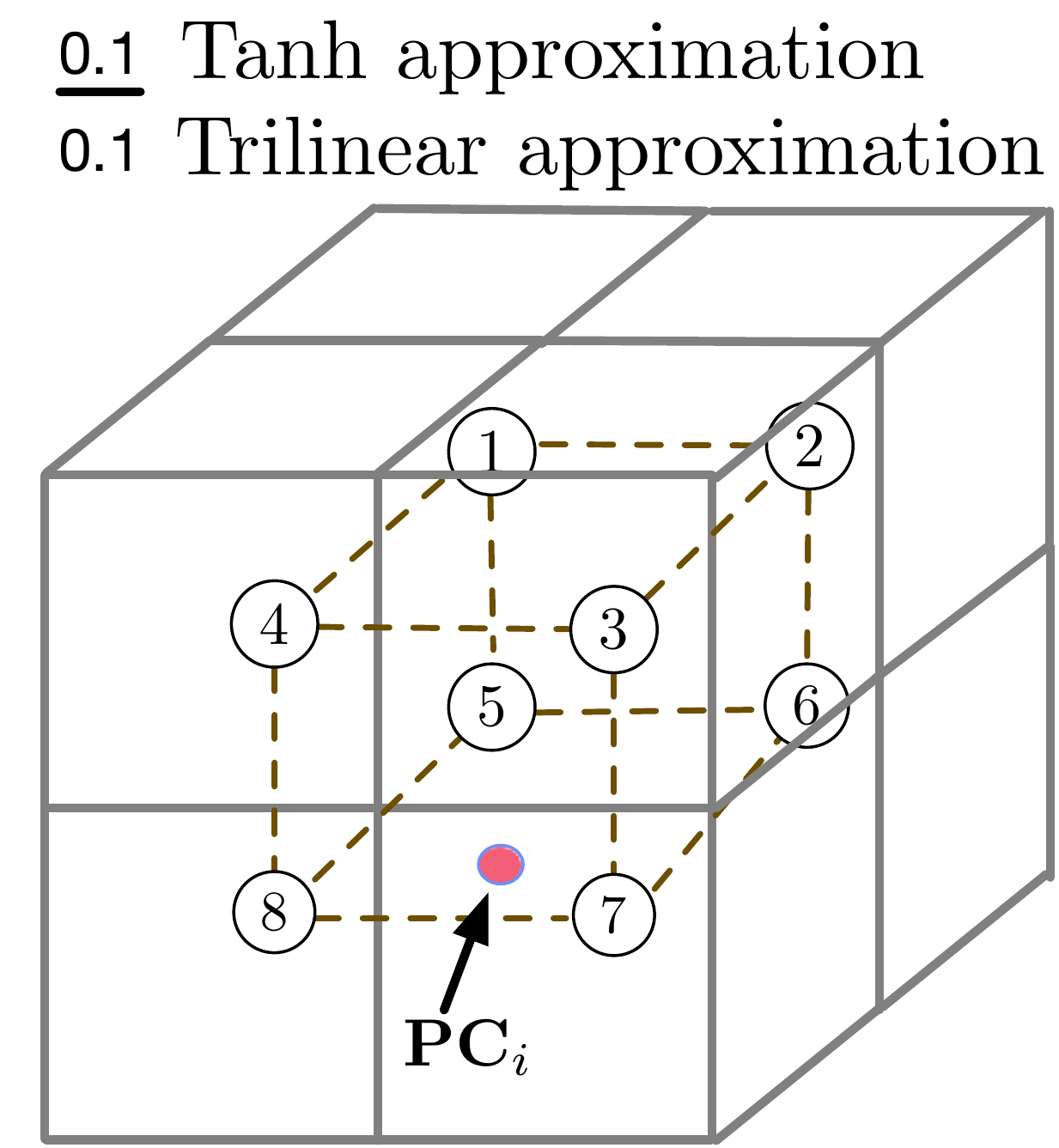}
         \vspace{-0.6cm}
         \caption{\scalebox{0.9}{8 cells \& formed cube}}
         \label{fig:8cells}
         \end{subfigure}
    \hspace{0.01\linewidth}
     \begin{subfigure}{0.32\linewidth}
         \centering
         \vspace{-0.09cm}
         \includegraphics[width=1.\linewidth]{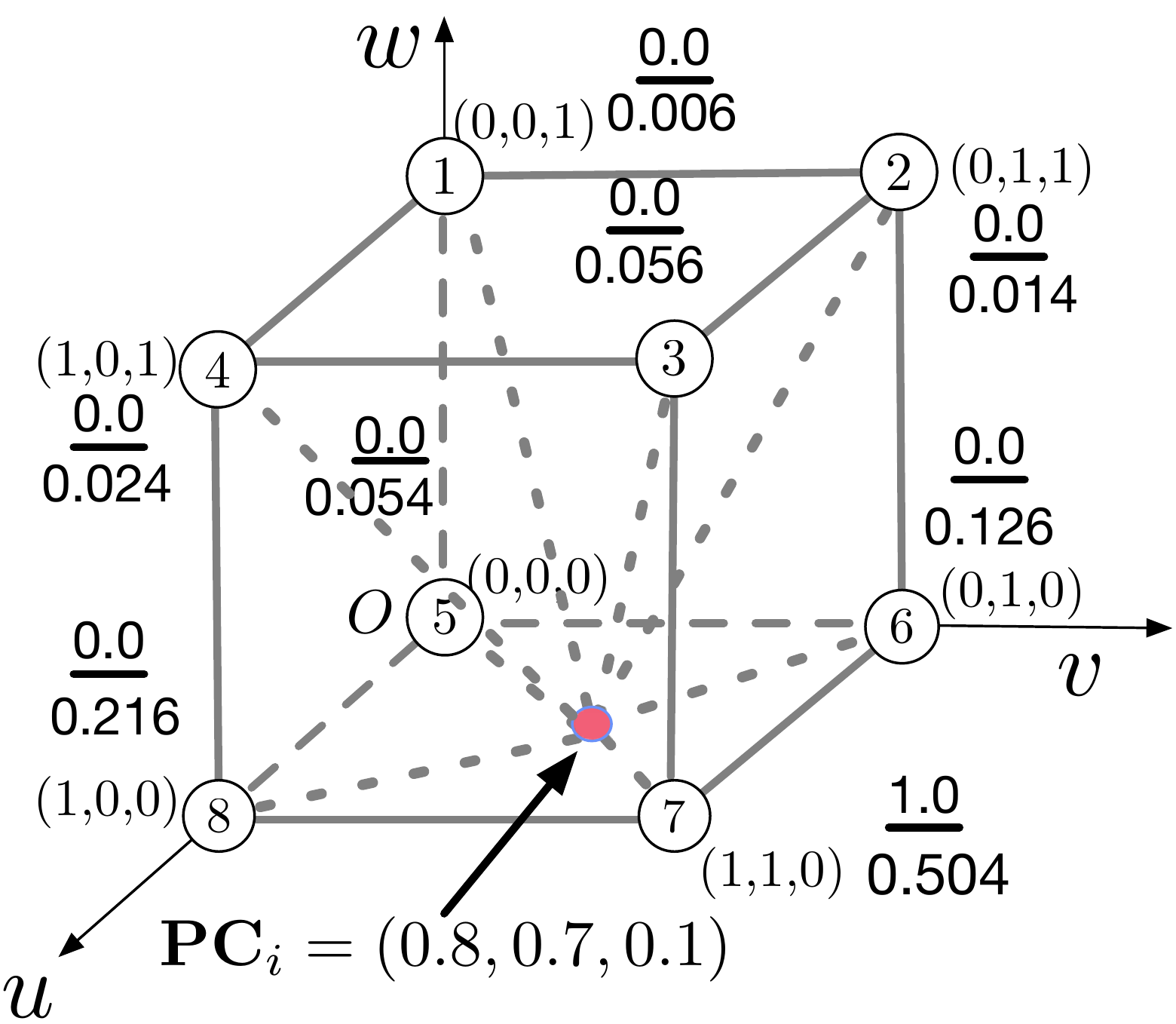}
         \vspace{-0.59cm}
         \caption{\scalebox{0.9}{Soft point-inclusion calc.}}
         \label{fig:formedcube}
     \end{subfigure}
    \hspace{0.01\linewidth}
     \begin{subfigure}{0.28\linewidth}
         \centering
        \vspace{-0.0cm}
         
         \includegraphics[width=1.\linewidth]{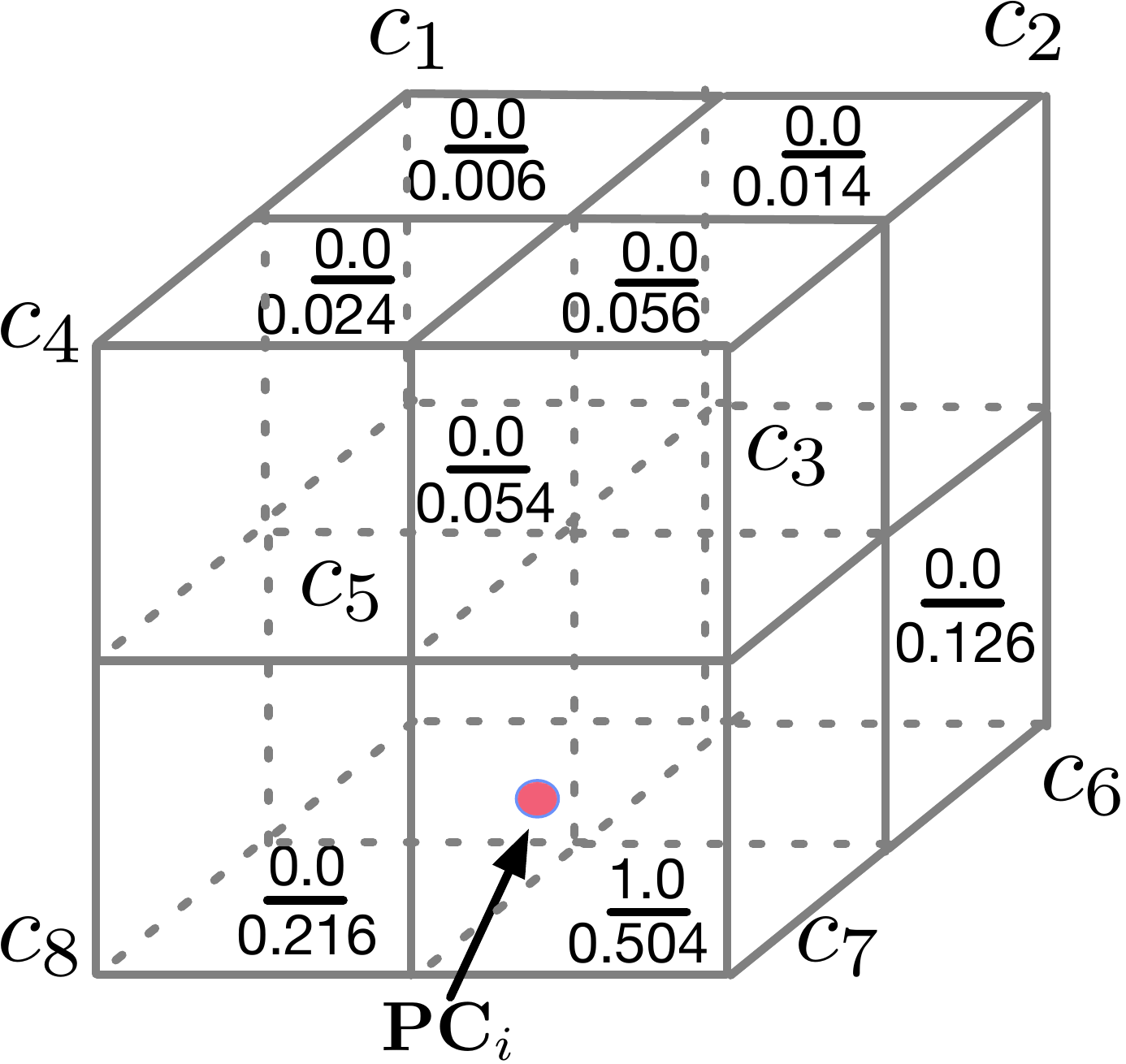}
         \vspace{-0.56cm}
         \caption{\scalebox{0.9}{Result assigned to 8 cells}}
         \label{fig:valueassign}
     \end{subfigure}
    \vspace{-0.1in}
    \caption{Illustration of the soft point-inclusion calculation with trilinear and tanh approximations. $\mathbf{PC}_i$ is a point in $\mathrm{PC}$, and $c_1$ to $c_8$ are the 8 3D cells closest to $\mathbf{PC}_i$.
    }
     
    \label{fig:ill-count}
    \end{minipage}
    \hspace{0.001\textwidth}
    \begin{minipage}[t]{0.3\textwidth}
        \footnotesize
          \centering
         
              \includegraphics[width=1.0\linewidth]{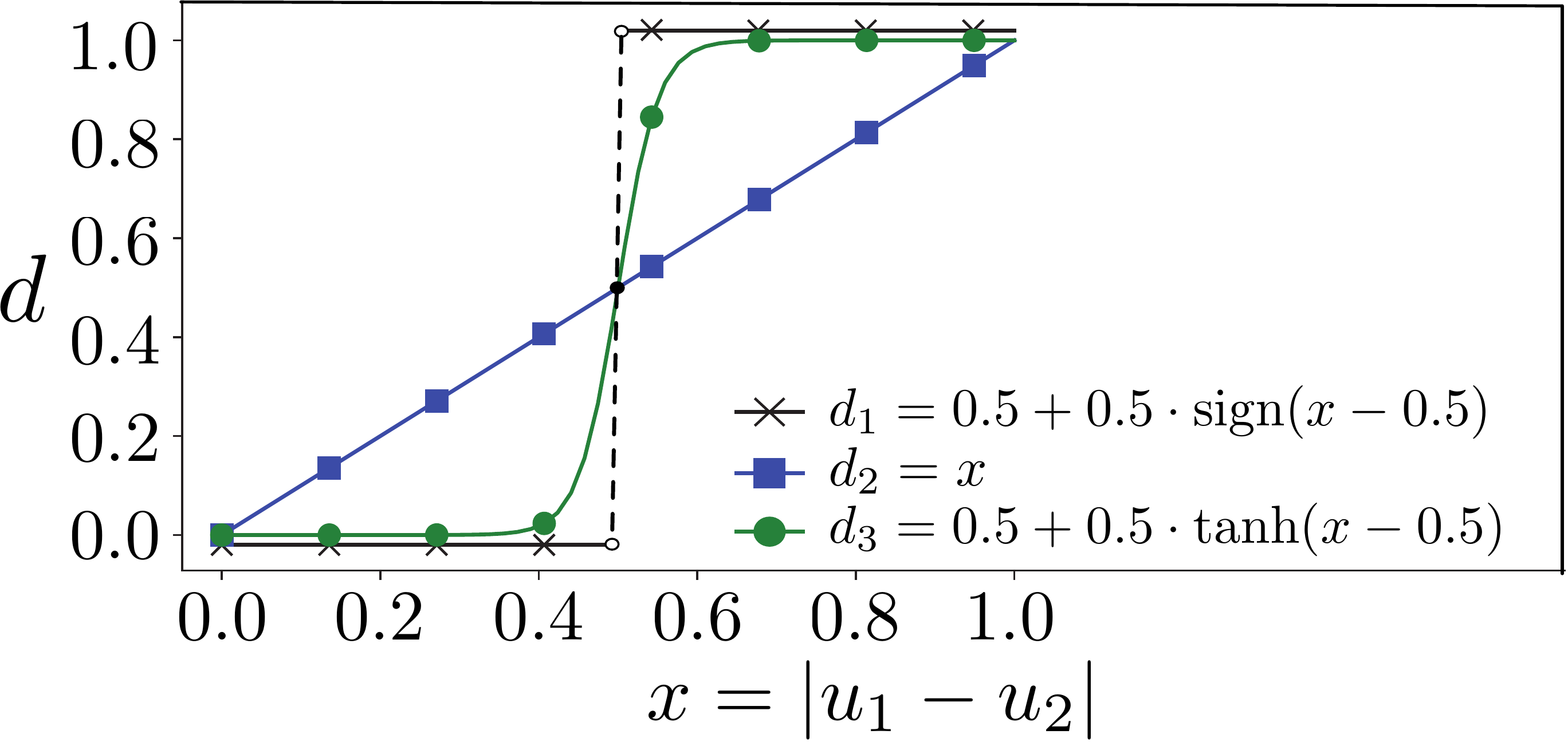}  
        \vspace{-0.43cm}
        
    	\caption{Line $d_1$ is the ground truth, while $d_2$ and $d_3$ are trilinear and tanh approximation. As shown, the tanh one is much closer to the ground truth.}
        \label{fig:tanherror}
    \end{minipage}

\end{figure*}

\textbf{Building block: Soft point-inclusion calculation.} Given a point $\mathbf{PC}_i$ with coordinate $(u_i, v_i, w_i)$ from the point cloud $\mathrm{PC}$ and a 3D cell $c_m$ of length $L$, width $W$, and height $H$, the direct point-inclusion value of $\mathbf{PC}_i$ for $c_m$, denoted as $\mathrm{PI}(\mathbf{PC}_i, c_m)$, is 1 if $\mathbf{PC}_i$ is inside $c_m$, and 0 if not. To differentiably approximate this function, we estimate the \textit{point-inclusion probability} of the point among the 8 cells closest to it by calculating the interpolation of it to these 8 cell center positions. Fig.\mref{fig:8cells} illustrates these 8 cells, which are indexed as $m=1...8$. The center position of a cell $c_m$ is denoted as $(u_m, v_m, w_m)$. These 8 center positions form a cuboid that encloses $\mathbf{PC}_i$. We can then calculate the interpolation of this point to these center positions using trilinear interpolation\mcite{trilinear}:

\begin{small}
\begin{equation} \label{eq:point-inclusion}
\begin{split}
    \mathrm{softPI}(\mathbf{PC}_i, c_m) = & (1 - \frac{d(u_m, u_i)}{L}) \cdot(1 - \frac{d(v_m , v_i)}{W}) \\
    & \cdot (1 - \frac{d(w_m, w_i)}{H})
\end{split}
\end{equation}
\end{small}

\noindent{}where $d(u_1, u_2) = |u_1 - u_2|$ and $\sum_{m=1}^8 \mathrm{softPI}(\mathrm{PC}_i, c_m) = 1$. Thus, this is similar to calculating the probabilities of whether $\mathrm{PC}_i$ is inside each of these 8 cells. Fig.\mref{fig:formedcube} illustrates the calculation process and the example calculation for $\mathbf{PC}_i$ at $\mathrm{(0.8, 0.7, 0.1)}$ when $L=W=H=1$ (i.e., each cell is a cube) and the center coordinate of $c_5$ is the origin $\mathrm{(0, 0, 0)}$. The calculation results are the numbers without underline at the 8 center positions. In Fig.\mref{fig:valueassign}, the interpolation value at the center position of each cell is then used as the point-inclusion probability for such cell. As shown, since $\mathbf{PC}_i$ is inside $c_7$, it is the closest to the center of $c_7$ at $\mathrm{(1, 1, 0)}$, and thus the interpolation value is the highest for $c_7$. This thus is able to correctly assign the highest point-inclusion probability to $c_7$.

\textbf{Approximation accuracy improvement.} In Fig.\mref{fig:formedcube}, while the point-inclusion probability is indeed the highest for $c_7$, the probability value is only 0.504 and thus still has a non-negligible gap to the ground-truth value 1. We find that the cause of this gap is at the $d(u_1, u_2)$ function in \meq{eq:point-inclusion}. As shown in Fig.\mref{fig:tanherror}, the ground-truth function for $d(u_1, u_2)$ when $L=W=H=1$ is $0.5+0.5\cdot{}\mathrm{sign}(|u_1-u_2|-0.5)$, since if the distance between the point and the cell center at any dimension is over 0.5, it is outside of cell and thus $(1-d(u_1, u_2))$ should be 0 in \meq{eq:point-inclusion}. Since $\mathrm{sign}(x)$ is not differentiable when $x=0$, such ground-truth function cannot be directly used in $\mathrm{softPI}(\cdot)$. Using $d(u_1, u_2) = |u_1 - u_2|$ as in classic trilinear interpolation is differentiable, but its curve has a gap to the ground truth as shown in Fig.\mref{fig:tanherror} so that it is more difficult for the optimized $S^a$ to succeed. To address this, we use $\mathrm{tanh}(\cdot)$ to differentiably and accurately approximate $\mathrm{sign}(\cdot)$. For example, for the $u$ dimension, it becomes:

\small
\begin{equation}
    \label{eq:tanh}
    d(u_1, u_2) = \frac{L}{2} + \frac{L}{2}\cdot\tanh(\mu\cdot(\mid u_1 - u_2 \mid - \ \frac{L}{2}))
\end{equation}
\normalsize

For the $v$ and $w$ dimensions of $\mathbf{PC}_i$ we replace $L$ with $W$ and $H$. Fig.\mref{fig:tanherror} shows the curve of \meq{eq:tanh} when $L=1$. As shown, the difference between \meq{eq:tanh} approximation and the ground truth is much smaller. In this paper, we call $\mathrm{SoftPI}(\cdot)$ using $d(u_1, u_2) = |u_1 - u_2|$ and \meq{eq:tanh} \textit{trilinear} and \textit{tanh approximation} respectively.  The numbers with underline in Fig.\mref{fig:formedcube} and (c) are the results with tanh approximation. As shown, with tanh approximation the point-inclusion probability for $c_7$ becomes 1.0, which is directly the ground-truth value and thus much more accurate than trilinear approximation.

\begin{figure}[t]
    \centering
     \includegraphics[width=\linewidth]{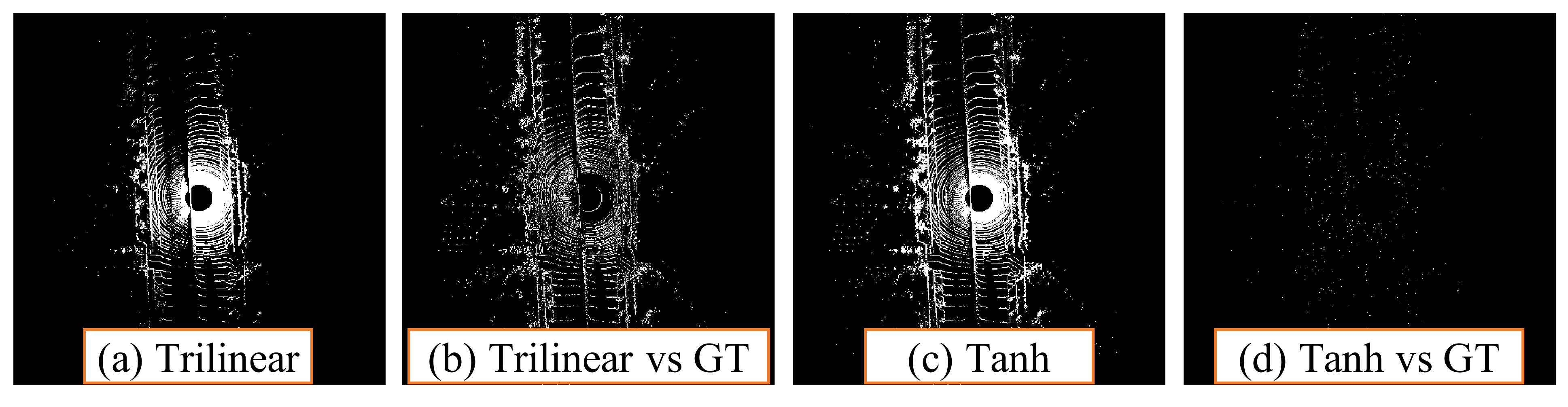}
    \caption{Accuracy benefit of $\mathrm{tanh}$ approximation over trilinear approximation for the count feature (number of points per cell). Count values are visualized using a gray-scale heatmap. GT denotes the ground-truth count value.}
    \label{fig:diff}
\end{figure}

To more concretely show the benefit of tanh approximation, Fig.\mref{fig:diff} shows the calculation results for the count feature in Table~\ref{tab:features-bev-voxel} based on $\mathrm{softPI}(\cdot)$ using real-world point cloud data. The count feature calculates the number of points in a cell (derivation of it from $\mathrm{softPI}(\cdot)$ is described later). In Fig.\mref{fig:diff}, the count values are visualized using a gray-scale heatmap in BEV. Fig.\mref{fig:diff} (a) and (c) shows the count values calculated using trilinear and $\mathrm{tanh}$ approximations respectively, and (b) and (d) shows their differences to the ground-truth count value. As shown, the count values using trilinear approximation have clear differences to the groundtruth, while the differences for the ones using tanh approximation is almost invisible.

\textbf{Derivation of cell-level aggregated features.}
With an accurate $\mathrm{SoftPI}(\cdot)$, we can then differentiably approximate all the cell-level aggregated features in Table~\ref{tab:features-bev-voxel} as follows:

\labelitemi\indent\textit{Count and density.} The count feature calculates the number of points in a cell. With $\mathrm{softPI}(\cdot)$, we can differentiably derive the count value as $\mathrm{CNT}(c_m) = \sum_{\mathbf{PC}_i \in \mathrm{PC}} {\mathrm{softPI}(\mathbf{PC}_i, c_m)}$. The density feature calculates the density of points in a cell. Thus, we can directly calculate it by dividing $\mathrm{CNT}(\cdot)$ by the cell size. 

\labelitemi\indent\textit{Occupancy.} The occupancy feature calculates whether a cell has points or not. With $\mathrm{CNT}(\cdot)$ above, it can be calculated as $\mathrm{sign(\mathrm{CNT}(\cdot))}$. Note that since the $\mathrm{sign}(\cdot)$ is not differentiable, we approximate it using $\sign(x) = x$ during the backward pass of the optimization.

\labelitemi\indent\textit{Height and intensity.} The max/min/mean height features calculate the maximum, minimum, and the average height of the points inside a cell. Thus, the max and min height features are directly $\max_{\mathbf{PC}_i \in \mathrm{PC}}{\mathrm{softPI}(\mathbf{PC}_i, c_m) \cdot w_i}$ and $\min_{\mathbf{PC}_i \in \mathrm{PC}}{\mathrm{softPI}(\mathbf{PC}_i, c_m) \cdot w_i}$. The mean height feature can be calculated as $\frac{\sum_{\mathbf{PC}_i \in \mathrm{PC}}{\mathrm{softPI}(\mathbf{PC}_i, c_m) \cdot w_i }}{\mathrm{CNT}(c_m) + \epsilon}$, where $\epsilon$ is small number to prevent division by zero. The max/min/mean intensity features can be calculated similarly by replacing $w_i$ with the intensity value of $\mathbf{PC}_i$.

The calculations above are performed for 3D cells. To obtain features for 2D cells, we just need to add an aggregation of these 3D cell features in one dimension, e.g., the vertical dimension for BEV 2D cells (\S\ref{sec:background-msf}), into these calculations.

\subsection{Objective Function Design}
\label{sec:objective-function}

\textbf{Adversarial loss $\gL_a$.} For the adversarial loss $\gL_a$ in \meq{eq:overview-opt}, similar to prior attacks on object detection\mcite{eykholt2018physical, zhao2018seeing}, 
we extract and minimize the confidence value (which reflects the confidence that the region contains an object) of the regions of $S^a$. 
As introduced in~\S\ref{sec:background-msf}, the fusion process of the LiDAR and camera perception networks in the MSF algorithm can be DNN-based or rule-based. For the former, we directly extract the confidence values in the MSF output\mcite{chen2017multi, xu2018pointfusion, frossard2018end, liang2018deep, liang2019multi, du2017car, ku2018joint}. For the latter, since the rule-based fusion logic is not directly differentiable, we extract the confidence values in the outputs of the LiDAR and camera perception networks separately, and minimize the sum of them. This is because if we can prevent $S^a$ from being detected in the outputs of both the LiDAR and camera perception networks, $S^a$ will not appear in the MSF output no matter what the rule-based logic is.

\textbf{Realizability loss $\gL_r(\cdot)$.} To realize our attack goal in~\S\ref{sec:attack-goal}, $S^a$ needs to be 3D-printed and placed on top of the road surface in the physical world. To facilitate this, we design the realizability loss $\gL_r(\cdot)$ in our objective function to (1) improve the printability of $S^a$ at 3D printers by maximizing its surface smoothness using a Laplacian loss\mcite{laplacian1}, and (2) prevent the generation of $S^a$ that is underneath the road surface. The detailed loss formulations are in Appendix~\ref{sec:realize-loss}.

\textbf{Stealthiness designs.} Our optimization process has two designs for improving the stealthiness of $S^a$. First, the realizability loss above can improve its surface smoothness, which can thus allow it to look normally in practice. Second, we solve \meq{eq:overview-pgd} by using Project Gradient Descent (PGD) with $L_\infty$ distance constraint during the gradient update step in Fig.\mref{fig:framework}, which thus ensures that the per-dimension moving distance for each vertex in $S$ is smaller than $\epsilon$. We can then use $\epsilon$ to control how similar $S^a$ looks compared to the benign one $S$, and thus the smaller $\epsilon$ is, the stealthier $S^a$ is.

\textbf{Attack robustness improvement.} To achieve the end-to-end attack success in our setting, it is ideal if $S^a$ can be continuously undetected by the MSF algorithm when the victim AV is approaching the object, until their distance is smaller than the brake distance\mcite{brake-distance} so that it is too late to brake to avoid the crash. Thus, we need to improve the robustness of $S^a$ against different victim approaching distances and angles of the target road. To achieve this, we implement Transformation $T$ via random yaw-dimension rotations and ground-plane position shifting of $S^a$, which is illustrated in Fig.\mref{fig:framework}.
\nsection{Attack Evaluation}
\label{sec:evaluation}

\vspace{0.05in}
\subsection{Evaluation Methodology and Setup}
\label{sec:eval_setup}

{\bf MSF algorithm selection.} 
In our evaluation, we target MSF algorithms included in open-source industry-grade AD systems to ensure high practicality and realism of our evaluation results. In particular, we select the ones included in 2 open-source full-stack AD systems, Baidu Apollo\mcite{apollo} and Autoware.AI\mcite{autoware}, due to their (1) \textit{representativeness} among industry-grade AD systems today, as Apollo has been recently ranked among the top 4 leading industrial AD developers along with Waymo, Ford, and Cruise\mcite{ranking-AD}, and Autoware is adopted by the USDOT in their AV fleet\mcite{carma-platform}; (2) \textit{practicality}, since both systems can be readily installed on real vehicle models\mcite{autoware-car,Apollo-lincon} for driving on public roads. In particular, Apollo has been providing self-driving taxi services in China for months\mcite{apollo-taxi}; and (3)  \textit{ease to experiment with}, since they are the only full-stack AD systems that are open-sourced.

Both AD systems use rule-based fusion in their MSF algorithms, i.e., the LiDAR and camera perception networks are separated DNN models, and their individual perception outputs are fused based on hard-coded matching and prioritization rules. As described in \S\ref{sec:background-msf}, such design has high modularity and is easy to debug, interpret, and hard-code safety rules/measures\mcite{modular-end2end}. These can greatly benefit system development in industry, which might be the reasons why it is adopted in both Apollo and Autoware.AI. As described in~\S\ref{sec:objective-function}, for such fusion type, our optimization objective is to make our adversarial object undetected in both the outputs of the LiDAR and camera perception models to allow our attack to succeed no matter what rule-based fusion logic is used.

Due to such modular fusion designs, the MSF algorithms in both Apollo and Autoware.AI allow different combinations of LiDAR and camera perception models. Thus, in our evaluation we also evaluate our attack against different such combinations to understand the generality of our attack. In this paper, we call each such combination an \textit{MSF combination} and use \textcircled{+} to denote such combination operation. In particular, we select 2 different models for LiDAR and 2 for camera, which forms 4 MSF combinations in total. On the LiDAR side, the LiDAR perception model in Apollo is also included in Autoware.AI. Thus, we choose 2 models in different Apollo versions that have substantially different DNN designs: one from the latest version, v5.5, denoted as \textit{A5-L}, and another from an older version, v2.5, denoted as \textit{A2-L}. At the DNN design level, \textit{A5-L} differs greatly from \textit{A2-L} with 43.9\% more deep layers and 65.0\% more trainable parameters. On the camera side, we select the one from the latest version of Apollo, denoted as \textit{A5-C}, and the pre-trained YOLO v3\mcite{yolo-darknet}, denote it as \textit{Y3}, which is included in the latest version of Autoware.AI. 

{\bf 3D object type selection.} Considering the supported object types for the LiDAR and camera models, we experiment with 3 types of objects for the above 4 MSF combinations: (1) a \textit{traffic cone} of size 0.5 m $\times$ 0.5 m $\times$ 1.0 m, for A5-L\textcircled{+}A5-C and A2-L\textcircled{+}A5-C, (2) a \textit{bench} of size 0.6 m $\times$ 0.5 m $\times$ 1.5 m, for A5-L\textcircled{+}Y3 and A2-L\textcircled{+}Y3, and (3) a \textit{toy car} of size 0.6 m $\times$ 0.7 m $\times$ 1.6 m (for kids to sit inside), for all 4 MSF combinations. We intentionally avoid large objects like cars since they are much harder to 3D-print and deploy. Among the 3 object types, we consider traffic cone as the most attractive for attacker since it is much more common to appear on the roadway than the other two and thus the most stealthy. Thus, majority of our experiments are focused on traffic cone. 

{\bf Attack scenario selection.}
For each object type, we select 100 real-world driving scenarios from the KITTI dataset\mcite{kittidataset} in which such object in benign case can be 100\% detected by the MSF combinations. Each scenario is one frame of sensor inputs including the camera image, the LiDAR point cloud, and the calibration matrix.
These scenarios has high diversity with different types of objects (e.g., cars, trucks, traffic lights) and roads (e.g., local, high-way, to rural roads).

{\bf Object placement.} For most experiments, we place the benign and adversarial objects  7 meters (m) in front of the victim. We choose 7 m because it is the braking distance\mcite{brake-distance} when the vehicle speed is 25 mph, almost the lowest one in normal driving. Since such distance is larger for higher vehicle speeds, 7 m represents the \textit{smallest} distance at which the object has to be detected by the victim to avoid a crash in normal driving scenarios. In \S\ref{sec:attack-robustness}, we also evaluate our attack among different victim distances and angles. More detailed attack parameter settings are in Table\mref{tab:para-setting} in Appendix.
\subsection{Attack Effectiveness}
\label{sec:effectiveness-attack}
In this section, we evaluate the effectiveness of our attack on the attack scenarios described in~\S\ref{sec:eval_setup}. 

{\bf Evaluation metrics.} Given an MSF combination and an attack scenario, we render our generated 3D adversarial object into the background point cloud and image, and test whether it can be detected by the MSF combination.
We determine our attack as success if and only if the adversarial 3D object is undetected by \textit{both} the LiDAR and camera models in such MSF combination. Under this criterion, the successful attacks can generally defeat \textit{any} rule-based fusion logic that can be applied to fuse the outputs of these two models. Thus, the calculated success rate is a \textit{lower bound} when a specific fusion logic is used, e.g., the ones in Apollo and Autoware.AI. We perform evaluation on 100 scenarios and report success rate.

\begin{table*}[t]
\footnotesize
    \centering
  \begin{tabular}{c|ccccccccc}
    \toprule
    
    \multicolumn{2}{c}{MSF Comb.}    & \multicolumn{2}{c}{A5-L\textcircled{+}A5-C } & \multicolumn{2}{c}{A5-L\textcircled{+}Y3} & \multicolumn{2}{c}{A2-L\textcircled{+}A5-C} & \multicolumn{2}{c}{A2-L\textcircled{+}Y3}\\
        \cmidrule(lr){3-4}
        \cmidrule(lr){5-6}
        \cmidrule(lr){7-8}
        \cmidrule(lr){9-10}
       
        \multicolumn{2}{c}{Object Type} & {Traffic cone} & {Toy car}& {Bench}& {Toy car} & {Traffic cone}& {Toy car} & {Bench}& {Toy car}\\
        \midrule

         \multicolumn{2}{c}{\multirow{-1}{*}{Success Rate}}  &
        $100\%$  &  $ 91\%$ & $100\%$ & $93\%$ &$100\%$ & $96\%$ & $100\%$& $97\%$\\
        \midrule
        &$\Delta \ell_1$& $5.92$ & $5.95$ &$5.93$& $5.97$ & $5.93$& $5.63$ & $5.90$& $5.61$\\
        &$\Delta \ell_2$&$3.28$ & $3.46$ &$3.39$& $3.37$&$3.43$ & $3.34$ & $3.30$& $3.25$\\
        \multirow{-3}{*}{\shortstack{Dist. \\ (cm)}}   &$\Delta \ell_\infty$ &
         $2.00$  & $2.00$  & $2.00$ & $2.00$ &$2.00$ &$2.00$ & $2.00$& $2.00$\\
        \midrule
          \multicolumn{2}{c}{LPIPS} & 0.06 & 0.02 & 0.20 & 0.04 & 0.07 & 0.17 & 0.20 & 0.06\\
        \bottomrule
        
    \end{tabular}
    \caption{Attack success rate and average vertex perturbation distance of \system on different MSF combinations in 100 driving scenarios. A5-L, A5-C: LiDAR and camera models in Baidu Apollo v5.5. A2-L: LiDAR model in Apollo v2.5. Y3: YOLO v3. All objects can be 100\% detected by each MSF combination in the benign case. }
    \label{tab:msfsuccrate}

\end{table*}

\begin{table}[t]
\centering
  \footnotesize
  \begin{tabular}{cccccc}
    \toprule
		&  &\multicolumn{3}{c}{$\Delta \ell_p$ Dist. (cm)} & \\
        \cmidrule(lr){3-5}
		\multirow{-2}{*}{\shortstack{Stealthiness \\ Level (cm)}}&  \multirow{-2}{*}{\shortstack{Succ.\\ Rate}} & $\Delta \ell_1$ & $\Delta \ell_2$& $\Delta \ell_\infty$ & \multirow{-2}{*}{LPIPS}\\
		
		\midrule
        $\epsilon = 2.0$ & $100\%$ & $5.92$ & $3.28$ & $2.00$ &0.06\\
		$\epsilon = 1.0$ & $93\%$  & $2.84$ & $1.51$& $1.00$ & 0.05\\
		$\epsilon = 0.5$ & $76\%$ &$1.38$ & $0.54$& $0.50$ & 0.05\\
\bottomrule
	\end{tabular}
	\caption{Stealthiness evaluation results of \system on MSF combination A5-L\textcircled{+}A5-C with the traffic cone object under different stealthiness levels of $\epsilon$ (\S\ref{sec:objective-function}).}
		\label{tab:Stealthiness}
\end{table}

{\bf Results.} The results for the 4 MSF combinations are shown in Table\mref{tab:msfsuccrate}. For all object types and all MSF combinations, success rates are at least 91\%, and those for traffic cone and bench are all 100\% among 100 driving scenarios. This shows that \system is an effective method.  
Among these results, the 100\% success rates for traffic cone is especially important, since it is the most attractive object type among the three from the attacker's view due to its small size and the ability to disguise as a normal traffic object in the middle of the road. Note that our method can achieve 91\% attack success rates even for the toy car of which the object type (car) has been heavily explored in training data of the model. 
Among the 4 MSF combinations, A5-L\textcircled{+}A5-C has the lowest attack success rates, which shows that the models from the latest version of Apollo are the most robust among the 4.

{\bf Stealthiness.} We also measure the stealthiness of our object using the average per-vertex $\Delta \ell_p$  distances and the LPIPS (Learned Perceptual Image Patch Similarity) metric~\cite{zhang2018perceptual}.
Table\mref{tab:msfsuccrate} shows the results with stealthiness parameter $\epsilon$ = 2 cm (\S\ref{sec:objective-function}). As shown, our attack only needs to move each vertex by 3.4 cm on average ($\Delta \ell_2$) to achieve at least 91\% success rates on all MSF combinations.
For LPIPS, we use the official implementation from~\cite{zhang2018perceptual} to measure the LPIPS value between the driving image with benign object rendered and the same image with the adversarial one rendered at the same location. As shown, the average LPIPS value is 0.10 across the 3 object types.
This is at the same level as those achieved in latest GAN-based image restoring methods~\cite{jo2020investigating}, which are generally considered as indistinguishable for human. In Table\mref{tab:Stealthiness}, we further evaluate our attack under different $\epsilon$ values on A5-L\textcircled{+}A5-C using traffic cone. As shown, the attack success rates are still over 93\% even when the average moved distance per vertex ($\Delta \ell_2$) is as small as 1.5 cm.

\textbf{Attack stealthiness user study.} To more directly evaluate the attack stealthiness, we also conduct a user study for traffic cone with 105 participants from Amazon Mechanical Turk~\cite{MTurk}. The results show that the generated adversarial traffic cone is generally viewed (1) as innocent as the original benign cone, and (2) less suspicious than certain benign ones with broken shapes. More details are in Appendix~\ref{sec:stealthiness-user-study}.

\textbf{Effectiveness under different attack settings.} We also perform evaluation under different attack parameter settings. We find that our attack is most sensitive to $\mu$, which show that the differentiable approximation design in~\S\ref{sec:preprocessing} is critical to the attack success. 
More details are in Appendix~\ref{sec:different-settings}.

{\bf Printability.} We also evaluate the printability of our attack using commercial printability checking tool and geometry metrics such as \textit{watertightness}\mcite{aaai3dobject,formlabs}, \textit{self-intersection}\mcite{gaussian-curvature}, and \textit{curvature}\mcite{gaussian-curvature}. Our results show that our generated objects are 100\% printable, and our printability improvement designs in \S\ref{sec:objective-function} substantially reduce the printing difficulties from 58.9\% to 74.3\%. 
Detailed are in Appendix~\ref{sec:attack-printability}.

{\bf Transferability.} We also evaluate the attack transferability among the 4 MSF combinations with the toy car object. We find that the transfer attack among them is generally effective, with success rates around 75\% on average.

\subsection{Comparison with Baseline Attack Methods}
\label{sec:eval-baseline}
While our attack shows high effectiveness in the previous section, it is unclear how much of it is due to the specific designs in \system. To understand this, in this section we compare our method with possible baseline attack methods.

{\bf Evaluation methodology.} We consider 2 baseline attack methods: (1) Gaussian noise based shape perturbation, denoted as \textit{GN}, and (2) Genetic algorithm~\cite{mitchell1998introduction} based attack generation, denoted as \textit{GA}. GN is used to understand whether the success of our attack is due to our optimization-based design (\S\ref{sec:methods}), or simply due to the nature of that level of shape perturbations. GA still uses our objective function design in~\S\ref{sec:objective-function} as fitness function, but does not need differentiability, which is thus used to understand whether our differentiable approximation function designs in~\S\ref{sec:preprocessing} are actually useful.

{\bf Experimental setup.} We perform comparison with our attack on A5-L\textcircled{+}A5-C MSF combination with the traffic cone object using the same setup in~\S\ref{sec:eval_setup}. We implement GN and GA using the corresponding standard Python libraries~\cite{gaussiannoise, geneticalgorithm}. For GN, we apply a Gaussian noise with $\mu = 0$ and $\sigma = 2.1$ cm to each vertex dimension to generate a similar level of perturbation as \system with $\epsilon$ = 2 cm. For GA, we set the population size to 50, a common value used in genetic algorithm based adversarial attacks~\cite{alzantot2019genattack, Feng2020CGATTACKMT}. We configure it to use 2 cm as the per-dimension perturbation bound for each vertex, the same as $\epsilon$ in \system. To achieve a fair comparison, we run GA using similar CPU and GPU resources as \system, and ensure that it runs longer than our method.

{\bf Result.}
Table~\ref{tab:baseline-result} summarizes the attack success rates of GN, GA, and our method, and the corresponding shape perturbation degrees. As shown, for GN, the average moved distance per vertex is 3.35 cm ($\Delta \ell_2$), which is larger than those generated by our method (3.28 cm). However, only 8\% of the ones from GN succeed, which is a magnitude lower than ours (100\%). This thus shows that our high attack effectiveness is mainly due to our optimization-based design, instead of the nature of a similar-level shape perturbation. For GA, we stop it after it generates 2000 adversarial objects for each attack scenario, which is twice the number for our method (1000). However, the success rate is only 9\%, which is also a magnitude lower than ours. Fig.~\ref{fig:baseline} shows the fitness value trend during the optimization process, which is averaged over the 100 attack scenarios. As shown, the fitness value decrease for GA is much slower than ours: its fitness value drop after 2000 trials is achieved after only 133 trials using our method, which is 15$\times$ more efficient. This thus concretely shows that benefit of our differentiable approximation function designs in~\S\ref{sec:preprocessing}, which allows the use of gradient-based optimizations to significantly improve both the attack efficiency and effectiveness.

\begin{table}[t]
\begin{minipage}[b]{0.48\linewidth}
\centering

\setlength\tabcolsep{1.7pt}
\vspace{0.3cm}
	\begin{tabular}{ccccc}
    \toprule
		&  &\multicolumn{3}{c}{$\Delta \ell_p$ Dist. (cm)} \\
		\cmidrule(lr){3-5}
		 \multirow{-2}{*}{\shortstack{Attack \\ Method}}&  \multirow{-2}{*}{\shortstack{Success \\ Rate}} & $\Delta \ell_1$ & $\Delta \ell_2$& $\Delta \ell_\infty$\\
		\midrule
        GN  & $8\%$ & 21.8& 3.35& 10.3\\
		GA  & $9\%$ & 2.85& 1.84&2.00 \\
		\textbf{Ours} & $100\%$ & $5.92$ & $3.28$ & $2.00$\\
\bottomrule
	\end{tabular}
    \caption{Comparison between \system and baseline attack methods in attack success rate and object perturbation degrees. GN: Gaussian noise. GA: genetic algorithm.}
    \label{tab:baseline-result}
\end{minipage}
\hfill
    \begin{minipage}[b]{0.49\linewidth}
    \centering
    \includegraphics[width=40mm]{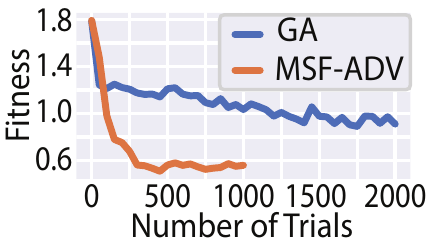}
    \vspace{-0.1cm}
    \captionof{figure}{Fitness values of GA and \system during the optimization process. The curve for \system stops at 1000 since it can already succeed for all scenarios at that point.}
    \label{fig:baseline}
\end{minipage}
\end{table}

\subsection{Attack Robustness}
\label{sec:attack-robustness}
In this section, we evaluate our attack robustness against different victim approaching positions and angles. 

{\bf Evaluation methodology.} We still use the attack scenarios in~\S\ref{sec:eval_setup} for  evaluation. To synthesize different relative positions between the victim and the object when the victim is approaching the object, we render the object at different locations ahead of the victim in both the camera and LiDAR frames given an attack scenario.

{\bf Experimental setup.} As described in our attack design (\S\ref{sec:design-overview}), the adversarial object is placed in the middle of the traffic lane in which the victim is driving. In this section, $X$ and $Y$ denote the relative distance between the victim and the object in the longitudinal (i.e., forward and backward) and lateral (i.e., left and right) directions respectively. For $X$, we consider 3 distance ranges from 5 to 35 m, which correspond to the brake distances for speed from $\sim$20 to 55 mph\mcite{brake-distance}.  For $Y$, the deviations to the center of the lane usually need to be within 0.1 m for smooth and safe driving\mcite{alika2020optimization-10cm-1, dominguez2016comparison-10cm-2}. Thus, we consider $Y \in (-0.1 \ \text{m}, 0.1 \ \text{m})$. For each position range, we randomly sample 20 different positions.

{\bf Results.}
Table\mref{tab:robustness} shows the average attack success rates for A5-L\textcircled{+}A5-C with traffic cone in the 3 position ranges over the 100 evaluation scenarios (\S\ref{sec:eval_setup}). As described in~\S\ref{sec:objective-function}, we use EoT to improve robustness. As shown, this improves the average success rates in all position ranges by 20.5\% on average.
Overall, with EoT the average attack success rates are over 95\% across different position ranges, 
which shows a high robustness of our attack against different victim approaching positions and angles at common driving speeds. 

\begin{table}[t]
\footnotesize
	\centering
	\setlength\tabcolsep{5pt}
  \begin{tabular}{cccc}
    \toprule
		& \multicolumn{3}{c}{Y = (-0.1 m, 0.1 m)
        }  \\ 
        \cmidrule(lr){2-4}
		& {X = (5 m, 15 m)} & {(15 m, 25 m)} & {(25 m, 35 m)} \\
\midrule
        w/o EoT & $80.3\%$& $79.2\%$& $79.9\%$\\
		w/ EoT & $96.3\%$& $95.5\%$& $96.6\%$\\
\bottomrule
	\end{tabular}
	\caption{Average success rate on A5-L\textcircled{+}A5-C with traffic cone in different victim approaching distance ranges.}
	\label{tab:robustness}
\end{table}

\subsection{Physical-World Attack Realizability Evaluation}
\label{sec:attack-realizability}

While the results in prior sections show high effectiveness and robustness of our attack, the experiments are performed by digitally rendering the objects into camera and LiDAR inputs. Thus, it is unclear whether such high effectiveness can still be achieved after the adversarial object is 3D-printed and placed in the physical world. Thus, in this section we evaluate such physical-world realizabilty of our attack.

\subsubsection{Real Vehicle based Experiments}
\label{sec:real-vehicle-attack-eval}

At the early stage of this project, we had access to a real vehicle equipped with a high-end Velodyne HDL-64E LiDAR, and used it to perform physical-world experiments for LiDAR models. Unfortunately, later we lost the access to it and only have such real vehicle based experiments for the LiDAR-side evaluation. In this section, we report these results for LiDAR side, and will detail in the next section the physical-world experiments for both LiDAR and camera using a miniature-scale experiment setup. 

\begin{figure}[t]
    \centering
    \begin{minipage}{0.5\textwidth}
    
    \begin{minipage}{0.40\linewidth}
     \begin{subfigure}{\linewidth}
     \centering
     \vspace{-0.07in}
     \includegraphics[width=1.\linewidth]{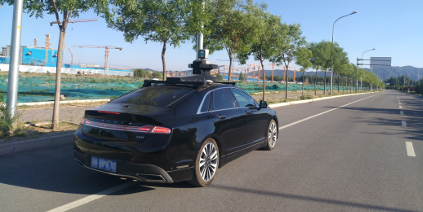}
     \vspace{-0.23in}
     \caption{\footnotesize Road \& car w/ LiDAR \label{fig:lidar-phy-settings}}
     \end{subfigure}
     \begin{subfigure}{\linewidth}
     \centering
     \vspace{0.03cm}
        
     \includegraphics[width=\linewidth]{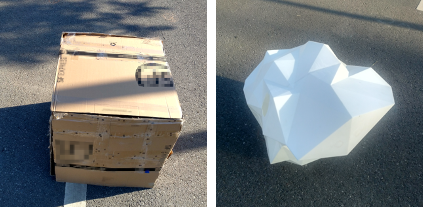}
     \caption{ Benign and adv. cubes  \label{fig:lidar-phy-ben} }
     
     \end{subfigure}
    \end{minipage}
     \begin{minipage}{0.6\linewidth}
     \vspace{-0.2cm}
     
      \begin{subfigure}{0.49\linewidth}
         \centering
         \includegraphics[width=1.\linewidth]{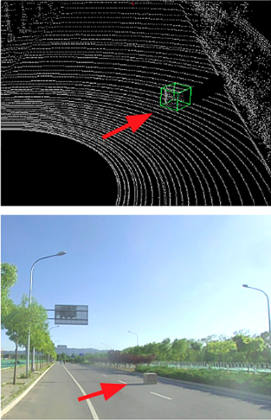}
         \vspace{-0.43cm}
         \caption{Benign case \label{fig:lidar-phy-ben-detection}}
         \end{subfigure}
         \hfill
         \begin{subfigure}{0.49\linewidth}
         \centering
         \includegraphics[width=\linewidth]{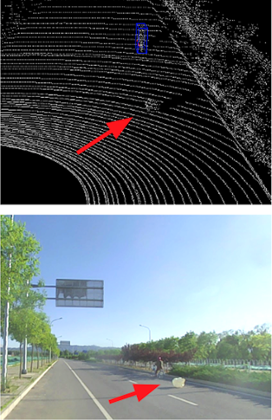}
         \vspace{-0.43cm}
         \caption{Adversarial case \label{fig:lidar-phy-adv-detection} }
         \end{subfigure}
     
    \end{minipage}
    \caption{Physical-world experiment settings and evaluation results for LiDAR-side physical-world attack realizability. We use a Velodyne HDL-64E LiDAR mounted on a real vehicle. The adversarial cube is 3D-printed at 1:1 scale.}
    \label{fig:physical-setups}
    \end{minipage}
    \label{fig:physical-LiDAR}
\end{figure}

{\bf Evaluation methodology and setup.} In this experiment, we 3D-print the adversarial object and conduct the experiment by using the vehicle mentioned above to collect its LiDAR point clouds on the real road. Fig.\mref{fig:lidar-phy-settings} shows the vehicle and road. We selected a rarely-used road and no other vehicles passed by during this experiment. Since this experiment was performed at the early stage of this project, the selected object type was a 75cm cube, and the targeted model was A2-L, the latest version of the Apollo LiDAR model at that time. Fig.\mref{fig:lidar-phy-ben} shows the box of the same size used as the benign cube, and the 3D-printed adversarial cube. This setup mimics the attack scenario by placing an adversarial rock-shaped object (\S\ref{sec:design-overview}).

{\bf Results.} We manually drive the vehicle around the cube and collect traces in front of it and on the left of it. In total, there are 99 LiDAR frames with the benign cube, and A2-L is able to correctly detect it in 84.8\% (84) frames. In comparison, we find that the adversarial cube is detected in only \textit{0.9\% (1)} of the 108 LiDAR frames including it. Fig.\mref{fig:lidar-phy-ben-detection} and Fig.\mref{fig:lidar-phy-adv-detection} show examples of the frames and detection results for the benign and adversarial cubes respectively. These results show that our attack is still effective in the physical-world setting for the LiDAR side of MSF. Experiment videos and images are at \DemoWeb.

\begin{figure*}[ht]
    \begin{minipage}[b]{0.30\linewidth}
    \centering
            \includegraphics[width=0.950\linewidth]{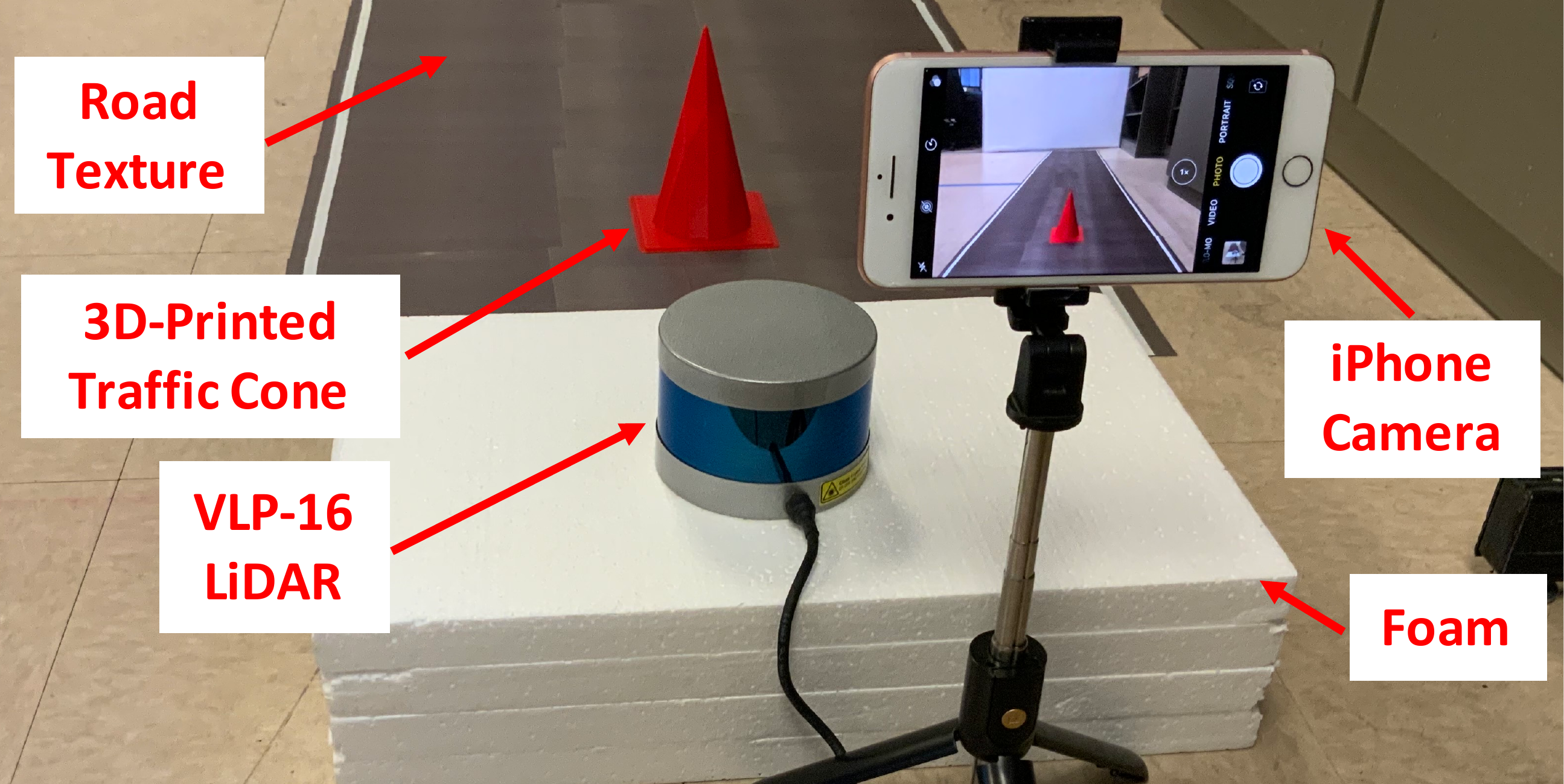}

        \caption{Miniature-scale experiment setup with camera and LiDAR. Road and traffic cone are at 1:6.67 scale.}
        \label{fig:miniature-physical-setting}
\end{minipage}
\hfill
\begin{minipage}[b]{0.683\linewidth}
\centering
            \includegraphics[width=\linewidth]{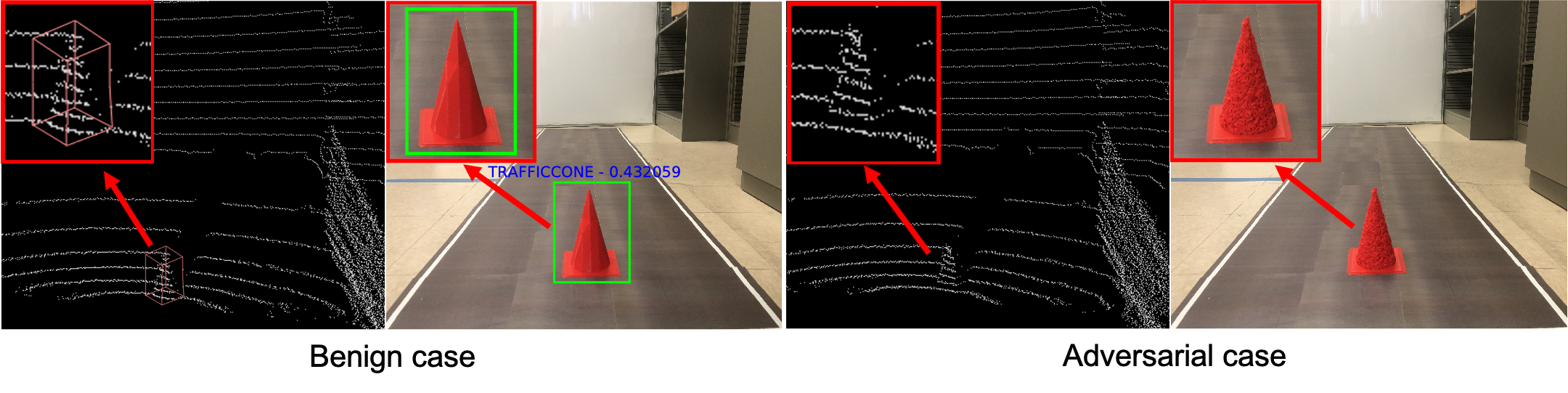}
            \vspace{-0.51cm}
        \caption{Visualization of the LiDAR and camera perception results for A5-L\textcircled{+}A5-C in miniature-scale experiments.}
        \label{fig:miniature-physical-res}
\end{minipage}
\end{figure*}

\subsubsection{Miniature-Scale Experiments}
\label{sec:miniature-scale-attack-eval}
Since we lost the access to the experiment vehicle, in this section we design a miniature-scale experiment in our lab environment to perform physical-world experiments for both LiDAR and camera.

\textbf{Evaluation methodology.} In this experiment,
we still 3D-print the adversarial object and obtain its point clouds and images using physical LiDAR and camera devices like in the actual physical-world attack settings. However, the main difference is that the adversarial object and the road are set up in a \textit{miniature scale} as shown in Fig.~\ref{fig:miniature-physical-setting}. As shown, the adversarial object is 3D-printed at 1:6.67 scale and placed on a miniature-scale straight road created by printing a real-world high-resolution BEV road texture on multiple A4 papers and concatenating them together. Here, the obtained point clouds of the object and road are scaled up accordingly following the physical rule of LiDAR to obtain the point clouds in real-world scale. The benefit of such miniature-scale setup is that it can not only obtain physical-world point clouds and images following the same physical rules of LiDAR and camera, but also more easily fit into the budget of a university-level research lab (e.g., 3D-printing our 1-meter high traffic cone at 1:1 scale requires industry-grade 3D printers~\cite{large-format-3d-print}).

\textbf{Experimental setup.} We use an iPhone 8 Plus back camera and a Velodyne VLP-16 LiDAR to collect images and point clouds as shown in Fig.~\ref{fig:miniature-physical-setting}. For the adversarial object, we generate the adversarial traffic cone mesh using the image and point cloud collected in our miniature-scale setup as the background.
We 3D-print the benign and adversarial traffic cones with 380 um precision at 1:6.67 scale. The road size, traffic cone size, and the camera and LiDAR positions are chosen to represent the scenario where these sensors are installed on a car driving on a standard 3.6-meter wide highway road~\cite{lanewidth}.

In the experiment, we try 20 different positions on the miniature road, which are randomly sampled in a 6.0 cm $\times$ 6.0 cm area at the road center and $\sim$45 cm far from the camera and LiDAR. We choose this area because we find the highest detection rate of the benign cone in this area. In real-world scale, this represents the scenario where the adversarial cone is roughly at the road center and 3-3.5 m far from the camera and LiDAR on the victim. Since the object type is traffic cone, we consider A5-C on camera side, and the VLP-16 versions of A5-L and A2-L in Apollo, which has the same model architecture as their corresponding HDL-64 versions~\cite{apollo-github}.

\begin{table}[t]{
    
	\centering
    \setlength\tabcolsep{3.5pt}
	\begin{tabular}{ccc}
        \toprule
        & { A5-L\textcircled{+}A5-C } & {A2-L\textcircled{+}A5-C (transfer attack)}\\
        \midrule
        Benign detection rate & 19/20 (95\%) & 16/20 (80\%)\\
        Attack success rate &18/20 (90\%) & 17/20 (85\%) \\
        Attack success rate when &  &\\
benign can be detected & \multirow{-2}{*}{18/19 (94.7\%)}& \multirow{-2}{*}{14/16 (87.5\%)}\\
\bottomrule
		
	\end{tabular}
	
	\caption{Evaluation results for A5-L\textcircled{+}A5-C and A2-L\textcircled{+}A5-C at 20 randomly-sampled positions in miniature-scale experiments. Results for A2-L\textcircled{+}A5-C is a transfer attack since the adversarial traffic cone is generated for A5-L\textcircled{+}A5-C.}
    \label{tab:miniature-result}
    
    }

\end{table}

\textbf{Results.}
Table~\ref{tab:miniature-result} shows the results. As shown, for A5-L\textcircled{+}A5-C, the benign traffic cone can achieve 95\% detection rate at the 20 random positions. However, after we place the adversarial one at exactly these 20 positions, the detection rate is only 10\%, leading to a 90\% success rate. Specifically, at the 19 positions that the benign cone can be successfully detected, the attack success rate is around 95\%. Fig.~\ref{fig:miniature-physical-res} visualizes the LiDAR and camera perception results of the benign and adversarial cones. More images and dynamic moving videos are at \DemoWeb.

Since this adversarial cone is generated for A5-L\textcircled{+}A5-C, we also evaluate it against A2-L\textcircled{+}A5-C to understand whether such attack effectiveness can transfer. As shown, the success rate of such a transfer attack is very similar: the success rate among the 20 positions is 85\%, and that among the positions where the benign cone can be detected is 87.5\%. These results thus show that our generated adversarial objects can still be effective against both LiDAR and camera in a physical-world environment, and such effectiveness can transfer. 

\nsection{End-to-End Attack Simulation Evaluation}
\label{sec:end2end}

To more concretely understand the end-to-end safety consequences, we further evaluate on a concrete attack scenario using a production-grade AD simulator. 

{\bf Evaluation methodology and metrics.}
We perform an end-to-end attack evaluation on Baidu Apollo using LGSVL simulator\mcite{LGSVL}.
LGSVL is an open-source Unity-based simulator designed for testing and development of industry-grade AD systems, and has already supported Apollo. 
In our evaluation, we use a map of a single-lane road in LGSVL, and set up Apollo to control a vehicle to drive along this lane.
To launch our attack, we imported the 3D mesh of our adversarial traffic cone into Unity, set its physical properties, and then re-build the simulator and the map.
We control the position of this adversarial cone to set it to the lane center, and LGSVL will provide Apollo with the raw camera and LiDAR inputs with the adversarial objects using its simulation engine.
As described in~\S\ref{sec:design-overview}, crashing into such an adversarial traffic cone can lead to severe safety damages as the attacker can fill it with denser materials such as granite or metal, or put nails or glass debris behind it.
Considering such concrete attack scenarios, we directly use the vehicle collision rate with the adversarial cone to evaluate the attack effectiveness.

{\bf Experimental setup.} We evaluate on Apollo v5.0, the latest Apollo version supported by LGSVL so far\mcite{LGSVL}. We use the default camera and LiDAR device configurations in this support. The LiDAR and camera models in Apollo v5.0 are the same as those in the latest version, Apollo v5.5. Thus, we directly use the adversarial traffic cone generated in \msec{sec:evaluation} for this evaluation. The vehicle speed is set to 30 km/h. For both benign and adversarial scenarios, we perform 100 runs of experiments and each lasts around 20 seconds to allow the vehicle to arrive at the traffic cone placement position and finish executing the driving decision. 

{\bf Results and demo videos.}
The results show that our adversarial traffic cone can \textit{always} fool the Apollo system in the entire trip across the 100 runs, leading to a \textit{100\%} vehicle collision rate. We inspect the experiment log and find that the adversarial cone evades both the camera and LiDAR perception pipelines at \textit{every} frames before fusion, which thus fundamentally defeats the basic design assumption of using MSF for defense. In contrast, in the benign case, Apollo is \textit{always} able to correctly detect the benign cone and stop in front of it to avoid collision (i.e., 0\% crash rate). Across different runs, the vehicle driving trajectories differ slightly due to the simulation randomness and sensor messaging delay/dropping, but our attack shows a high robustness against such trajectory variances when the victim is approaching.

\begin{figure}[t!]
      \centering
          \includegraphics[width=0.9\linewidth]{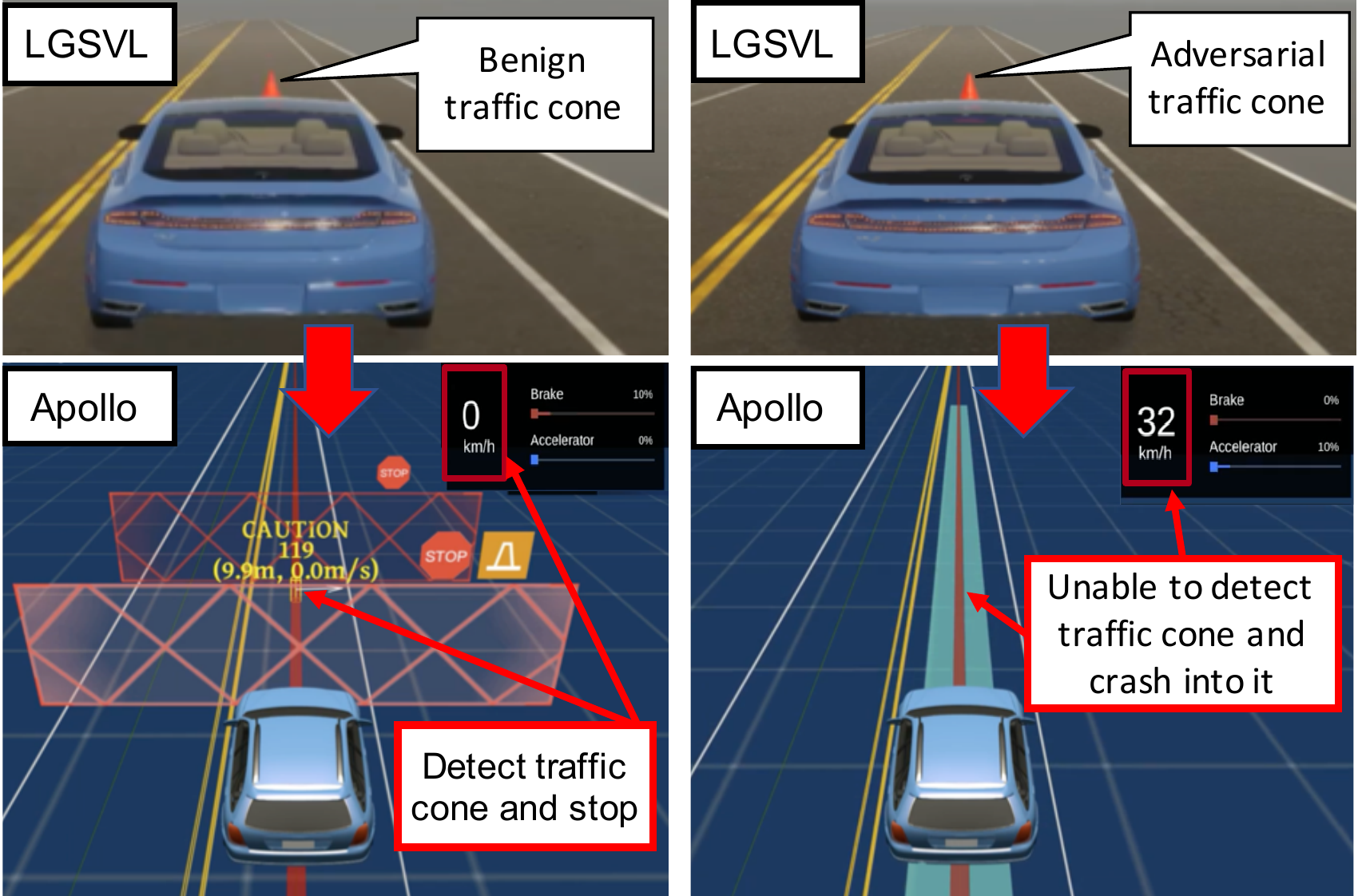}  
    \caption{Screenshots of Apollo and LGSVL in the end-to-end attack evaluation with benign and adversarial traffic cones. Across 100 runs, the crash rate is 100\% for adversarial case, and 0\% for benign case.}
        \label{fig:end-to-end-demo}
\end{figure}

Fig.\mref{fig:end-to-end-demo} shows the key screenshots on both the LGSVL and Apollo sides during the simulation. As shown, in the benign case, the victim can detect the traffic cone and successfully make a stop decision to decrease its speed to 0 km/h. However, in the adversarial case, the victim cannot detect the traffic cone even when it is right in front of it. Thus, it maintains the original speed and directly crashes into it. We also record short demo videos from the simulation, available at~\cite{ourwebsite}.

\section{Limitations and Defense Discussion}
\label{sec:discussion}
\subsection{Limitations of Our Study}
{\bf End-to-end physical-world evaluation.} In this work, our attack is designed with a practical attack model (\S\ref{sec:design-overview}) and evaluated on real-world driving dataset and miniature-scale physical-world settings (\S\ref{sec:attack-realizability}). However, we did not perform an end-to-end attack evaluation on a real AV in the physical world due to the cost and safety considerations. As a best effort, we evaluate such end-to-end attack impacts using a production-grade AD simulator (\msec{sec:end2end}). Note that AD companies such as Waymo also heavily rely on simulation-based evaluations when developing and testing AD systems for safety and budget considerations\mcite{company-simulation2}.

{\bf Attack generality evaluation.}
In our evaluation, we target the MSF algorithms used in representative industry-grade AD systems such as Baidu Apollo\mcite{apollo}, which generally adopt a rule-based fusion design. As introduced in~\S\ref{sec:background-msf}, there also exists another type of fusion design: DNN-based fusion\mcite{chen2017multi, xu2018pointfusion, frossard2018end, liang2018deep, liang2019multi, du2017car, ku2018joint}. Thus, it is still unclear how effective \system can be for DNN-based MSF algorithms. Note that this is not a limitation of our attack methodology: as described in \msec{sec:objective-function}, our design is generally applicable to both fusion designs. Also, since rule-based fusion design is more preferable for the system development in the industry (\S\ref{sec:eval_setup}), our current evaluation results can potentially lead to more impacts to AD systems in practice. Thus, we left the evaluation of MSF algorithms with DNN-based fusion as future work.

\subsection{Defense Discussion}
\label{sec:defense-discussion}

\subsubsection{DNN-Level Defense}
\label{sec:defense-dnn}

\begin{table*}[ht]
    \begin{minipage}[b]{0.73\linewidth}
    \centering
    \includegraphics[width=1.0\linewidth]{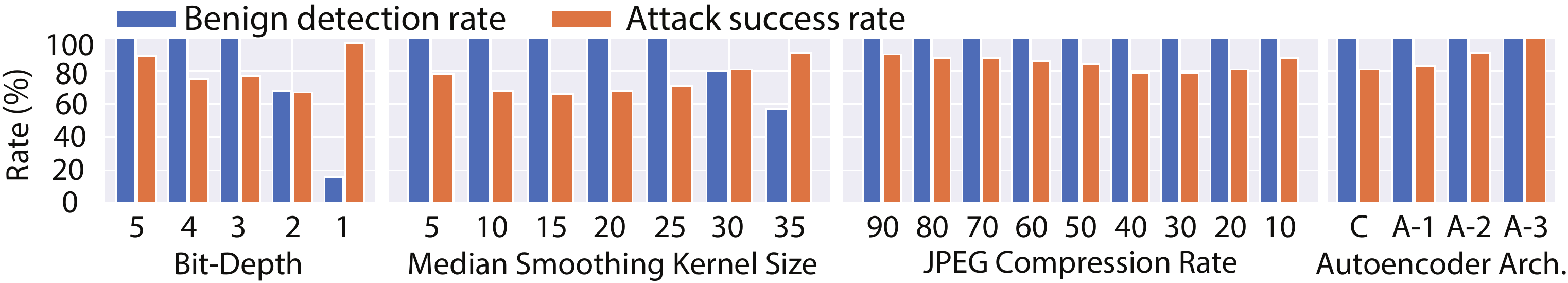}
    \vspace{-0.2cm}
    \captionof{figure}{Evaluation results of 4 DNN input transformation based defense methods for our attack on A5-L\textcircled{+}A5-C with traffic cone object. Benign detection rates mean detected by either LiDAR or camera. Attack success rate means that both LiDAR and camera fail to detect. For all x-axes, values from left to right mean higher to lower camera/LiDAR input quality (e.g., more smoothing or compression). Detailed setup in Appendix~\ref{sec:defense-setup}.}
    \label{fig:input-defense}
\end{minipage}
\hfill
\begin{minipage}[b]{0.26\linewidth}
\centering

\setlength\tabcolsep{2.0pt}
	\begin{tabular}{cccc}
	\toprule
        & Y3 test & Benign  & Attack\\ 
        & set mAP & det. rate  & succ. rate \\ 
        \midrule
        Original & 44\% & 100\%  &100\%\\
        AUG  & 32\%  &100\% & 69\% \\
    \bottomrule
	\end{tabular}
	\vspace{0.2cm}
	
	\caption{Results of augmenting training data (AUG)
	for our attack on A5-L\textcircled{+}Y3 compared to original model for bench object. Detailed setup in Appendix~\ref{sec:defense-setup}.}
    \label{tab:AT-AT}
\end{minipage}
\end{table*}

Our attack exploits vulnerability in DNNs used in MSF, and thus a direct defense direction is to secure these DNNs. In the recent arms race between adversarial attacks and defenses, various defense/mitigation techniques have been proposed, e.g., input transformation~\cite{Xu:2018:ndss, meng2017magnet, dziugaite2016study}, adversarial training~\cite{madry:towards}, and certified robustness~\cite{li2020sok, lecuyer2019certified}. However, almost all of them focus on image classification models under digital-space attacks, instead of object detection models under physical-world attacks. To the best of our knowledge, no prior works has considered defending against adversarial 3D objects in MSF context.  

\textbf{Experiment methodology.} In this case, as a best effort to understand the effectiveness of existing defenses in our attack setting, we perform experiments mainly on two easily-adaptable defense strategies: (1) camera/LiDAR input transformation without model re-training, for which we evaluate 4 popular methods: bit-depth reduction~\cite{Xu:2018:ndss}, median smoothing~\cite{Xu:2018:ndss}, JPEG compression~\cite{dziugaite2016study}, and autoencoder reformation~\cite{meng2017magnet}; and (2) augmenting training data, denoted as \textit{AUG}, which re-trains the model with adversarial inputs mixed in training dataset~\cite{szegedy:iclr:2014, goodfellow:fsgm, pei2017deepxplore}. AUG is only applied to YOLO v3 (Y3) since Apollo does not release training dataset for its models. Additionally, we also explored adversarial training~\cite{zhang2019towards} for Y3, but different from the standard adversarial training, we only applied {\em 2 steps PGD attack} to approximate the solutions of inner maximal problem for efficiency due to the complexity of our attack pipeline (e.g., rendering, pre-processing, and attacking two models together) caused by 
our problem settings and evaluated system. Such a strategy has been adopted in some recent works to improve efficiency of adversarial training~\cite{wong2020fast,zhang2019towards}. More details are in Appendix~\ref{sec:defense-setup}.
Note that we do not evaluate certified robustness~\cite{li2020sok, lecuyer2019certified} since its designs today focus on small 2D digital-space perturbations (e.g. $\ell_2$=0.5 on ImageNet~\cite{cohen2019certified, yang2020randomized}), and their extensions to either 3D space or physical-world attacks are still open research problems.

{\bf Results.}
Fig.~\ref{fig:input-defense} shows the results for the 4 input transformation based defenses on our attack on A5-L\textcircled{+}A5-C for traffic cone. For each method, we explore different parameters to explore the trade-off between benign detection rate and attack success rate.  As shown, with the decrease of the LiDAR/camera input quality (left to right for all the x-axes), the attack success rate will eventually increase for all 4 methods since the input quality becomes so low that both camera and LiDAR models cannot detect the object even in the benign case. For some methods, the attack success rate first decreases before such increase, which is likely because the input quality reduction disrupts our adversarial shape perturbations. Overall, median smoothing achieves the highest defense effectiveness by decreasing the attack success rate to 66\% without affecting the benign detection rate. Note that it is known that all these methods can be bypassed by adaptive attacks~\cite{carlini2017magnet, zhang2020interpretable, he2017adversarial, sharma2018bypassing}.
Thus, an interesting future work is to explore the effectiveness of these methods under adaptive attack designs of \system.

Table~\ref{tab:AT-AT} shows the results for AUG. For a fair comparison, the original model in the table is also newly-trained using the same setup. 
As shown, AUG is able to decrease the attack success rate to 69\% with 100\% benign detection rate. Our preliminary exploration of adversarial training with 2-step PGD does not show higher effectiveness: even with 900 epoch of training, the attack success rate is only reduced to 95\% with 100\% benign detection rate. The potential reason of the lower effectiveness is that 2 steps PGD is not enough to generate effective adversarial objects during training. Compared to some prior works~\cite{zhang2019towards, wong2020fast}, 
this suggests that our attack poses more challenges in balancing the trade-off between efficiency and effectiveness in adversarial training.
We plan to systematically investigate this in the future.

Overall, the most effective defense found in these experiments can only decrease the attack success rate to 66\%, which is not quite enough to render this attack vector practically unexploitable. Leveraging the analysis insights, we plan to explore more effective defense designs by exploring  (1) other input transformation considering the success of medium smoothing, and (2) more efficient and effective adversarial training designs for our attack. As certified robustness can provide strong theoretical guarantees, we also plan to explore the extensions of it to 3D space and physical-world attacks.

\subsubsection{Fuse More Perception Sources}
\label{sec:defense-fusion}

At MSF algorithm level, one defense direction is to fuse more perception sources, e.g., more cameras/LiDARs sharing an overlapped view but mounted at different positions, assuming that our attack may be more difficult to optimize if the fused camera/LiDAR perception results are from very different viewing angles and positions. Also, we may consider including RADAR into MSF, which is less preferred in state-of-the-art MSF designs (\S\ref{sec:background-msf}) but may help improve their security. Note that this cannot fundamentally defeat our attack since RADAR point clouds may also be affected by shape manipulations and their state-of-the-art object detection algorithms are still DNN-based\mcite{wang2019study}. Nevertheless, including RADAR may make it more difficult to attack if the RADAR perception model is more robust. We leave a systematic exploration of these to future work.
\section{Related Work}
\label{sec:literature}

\textbf{Autonomous Driving (AD) system security.} Since AD systems heavily rely on sensors, prior works have studied \textit{sensor attacks} in AD context such as spoofing/jamming attacks on camera~\cite{yan2016can, nassi2020phantom}, LiDAR~\cite{ shin2017illusion, cao2019adversarial}, RADAR~\cite{yan2016can}, ultrasonic~\cite{yan2016can}, and IMU~\cite{tu2018injected}. In comparison, these works mainly focus on vulnerabilities at sensor level, while we focus on those at the higher \textit{autonomy software} level, i.e., the ``brain'' of AD systems. At such level, prior works have studied the security of camera/LiDAR object detection~\cite{eykholt2018physical, chen2018shapeshifter, zhao2018seeing, cao2019adversarial, povolny2020adas} and tracking~\cite{jia2019fooling}, localization~\cite{junjie:usenix:2020}, lane detection~\cite{sato2020hold, sato2021wip, liang2021wip}, traffic light detection~\cite{kanglan2021}, and end-to-end AD~\cite{pei2017deepxplore, tian2018deeptest}. However, so far \textit{all} of them only consider attacks on camera or LiDAR perception \textit{alone}, while we are the first to study the security of MSF-based AD perception and address the corresponding design challenges (\S\ref{sec:design-challenge}).

\textbf{Adversarial attacks.} Various adversarial attacks have been proposed to generate adversarial attacks in the digital space\mcite{goodfellow:fsgm,carlini:cw,papernot:jsma,moosavi2016deepfool,szegedy:iclr:2014,pei2017deepxplore, xiao2019meshadv,aaai3dobject,xiang2019generating,lee2020shapeadv,xiao2018characterizing,qiu2020semanticadv,xiao2019characterizing,xiao2018spatially,xiao2018generating}. In comparison, we focus on physical-world attack vectors. Multiple prior works have designed and evaluated adversarial attacks in the physical world\mcite{eykholt2018physical, li2019adversarial,adv-patch,thys2019fooling,zhangcamou,chen2018shapeshifter,lu2017adversarial,zhao2018seeing,athalye2017synthesizing}. However, none of them have considered MSF-based AD perception, and as described in \msec{sec:design-challenge}, blindly combining their designs cannot directly achieve our goal due to various unique design challenges.

\section{Conclusion}
\label{sec:conclusion}

This paper presents a first study on the security issues of MSF-based AD perception, that challenges the basic design assumption for MSF as a defense strategy in AD context. We design a novel attack method, \system, with adversarial 3D object as the attack vector, and address design challenges in non-differentiable target camera and LiDAR sensing systems and non-differentiable computation of cell-level aggregated features for LiDAR. We perform evaluations on MSF algorithms included in industry-grade AD systems using real-world driving scenarios. Our results show that our attack achieves over 90\% success rates across different object types and MSF algorithms, while being stealthy, robust, transferable and physical-world realizable. In simulation evaluation, our attack can cause 100\% vehicle collision rate. We also evaluate and discuss defenses. Considering the critical role of perception for safe AV driving, we hope that our findings and insights can help the community develop effective defenses in practice.

\section*{Acknowldgments}

We would like to thank Nicolas Papernot, Junjie Shen, Ziwen Wan, Takami Sato, Junze Liu, Xinyang Zhang, Xi Lu, Yu Stephanie Sun, Joshua Garcia, Pengchuan Zhang, Hongge Chen, Dan Luo, Benjamin Emerson Dolan, and the anonymous reviewers for valuable feedback on our work. This research was supported in part by the NSF under grants CNS-1850533, CNS-1929771, CNS-1932464, and CNS-2012001, USDOT under Grant 69A3552047138 for the CARMEN UTC, the ARO under contract W911NF1810208, and grant from Open Philanthropy and Good Ventures Foundation.


\bibliographystyle{IEEEtran}
\bibliography{./main.bib}

\begin{thebibliography}{100}
\providecommand{\url}[1]{#1}
\csname url@samestyle\endcsname
\providecommand{\newblock}{\relax}
\providecommand{\bibinfo}[2]{#2}
\providecommand{\BIBentrySTDinterwordspacing}{\spaceskip=0pt\relax}
\providecommand{\BIBentryALTinterwordstretchfactor}{4}
\providecommand{\BIBentryALTinterwordspacing}{\spaceskip=\fontdimen2\font plus
\BIBentryALTinterwordstretchfactor\fontdimen3\font minus
  \fontdimen4\font\relax}
\providecommand{\BIBforeignlanguage}[2]{{%
\expandafter\ifx\csname l@#1\endcsname\relax
\typeout{** WARNING: IEEEtran.bst: No hyphenation pattern has been}%
\typeout{** loaded for the language `#1'. Using the pattern for}%
\typeout{** the default language instead.}%
\else
\language=\csname l@#1\endcsname
\fi
#2}}
\providecommand{\BIBdecl}{\relax}
\BIBdecl

\bibitem{sae2018}
S.~O.-R. A. V.~S. Committee \emph{et~al.}, ``{Taxonomy and Definitions for
  Terms Related to Driving Automation Systems for On-Road Motor Vehicles},''
  \emph{SAE International: Warrendale, PA, USA}, 2018.

\bibitem{av-companies}
``{40+ Corporations Working On Autonomous Vehicles},''
  \url{https://www.cbinsights.com/research/autonomous-driverless-vehicles-corporations-list}.

\bibitem{waymo-one}
``{Waymo has launched its commercial self-driving service in Phoenix - and it's
  called `Waymo One'},''
  \url{https://www.businessinsider.com/waymo-one-driverless-car-service-launches-in-phoenix-arizona-2018-12}.

\bibitem{tusimple-truck}
``{UPS joins race for future of delivery services by investing in self-driving
  trucks},''
  \url{https://abcnews.go.com/Business/ups-joins-race-future-delivery-services-investing-driving/story?id=65014414}.

\bibitem{eykholt2018physical}
K.~Eykholt, I.~Evtimov, E.~Fernandes, B.~Li, A.~Rahmati, F.~Tramer, A.~Prakash,
  T.~Kohno, and D.~Song, ``{Physical Adversarial Examples for Object
  Detectors},'' in \emph{WOOT}, 2018.

\bibitem{zhao2018seeing}
Y.~Zhao, H.~Zhu, R.~Liang, Q.~Shen, S.~Zhang, and K.~Chen, ``{Seeing isn't
  Believing: Practical Adversarial Attack Against Object Detectors},''
  \emph{ACM CCS}, 2019.

\bibitem{lu2017adversarial}
J.~Lu, H.~Sibai, and E.~Fabry, ``{Adversarial Examples that Fool Detectors},''
  \emph{arXiv preprint arXiv:1712.02494}, 2017.

\bibitem{zhangcamou}
Y.~Zhang, P.~D. Hassan~Foroosh, and B.~Gong, ``{CAMOU: Learning A Vehicle
  Camouflage For Physical Adversarial Attack On Object Detections In The
  Wild},'' in \emph{ICLR}, 2019.

\bibitem{chen2018shapeshifter}
S.-T. Chen, C.~Cornelius, J.~Martin, and D.~H.~P. Chau, ``{Shapeshifter: Robust
  Physical Adversarial Attack on Faster R-CNN Object Detector},'' in \emph{ECML
  PKDD}, 2018.

\bibitem{cao2019adversarial}
Y.~Cao, C.~Xiao, B.~Cyr, Y.~Zhou, W.~Park, S.~Rampazzi, Q.~A. Chen, K.~Fu, and
  Z.~M. Mao, ``{Adversarial Sensor Attack on LiDAR-based Perception in
  Autonomous Driving},'' in \emph{ACM CCS}, 2019.

\bibitem{jiachen:usenix:2020}
J.~Sun, Y.~Cao, Q.~A. Chen, and Z.~M. Mao, ``{{Towards Robust LiDAR-based
  Perception in Autonomous Driving: General Black-box Adversarial Sensor Attack
  and Countermeasures}},'' in \emph{Usenix Security}, 2020.

\bibitem{pei2017deepxplore}
K.~Pei, Y.~Cao, J.~Yang, and S.~Jana, ``{Deepxplore: Automated Whitebox Testing
  of Deep Learning Systems},'' in \emph{SOSP}, 2017, pp. 1--18.

\bibitem{tian2018deeptest}
Y.~Tian, K.~Pei, S.~Jana, and B.~Ray, ``{Deeptest: Automated Testing of
  Deep-Neural-Network-Driven Autonomous Cars},'' in \emph{ICSE}, 2018.

\bibitem{waymo-msf}
``{Waymo Tech},'' \url{https://waymo.com/tech/}.

\bibitem{apollo}
``{Baidu Apollo},'' \url{http://apollo.auto}.

\bibitem{autoware}
``{Autoware.AI},'' \url{https://www.autoware.ai//}.

\bibitem{ponyai-msf}
``{Pony.ai Tech},'' \url{https://www.pony.ai/en/tech.html}.

\bibitem{frossard2018end}
D.~Frossard and R.~Urtasun, ``{End-to-end Learning of Multi-sensor 3D Tracking
  by Detection},'' in \emph{ICRA 2018}.\hskip 1em plus 0.5em minus 0.4em\relax
  IEEE, 2018, pp. 635--642.

\bibitem{liang2018deep}
M.~Liang, B.~Yang, S.~Wang, and R.~Urtasun, ``{Deep Continuous Fusion for
  Multi-Sensor 3D Object Detection},'' in \emph{ECCV}, 2018.

\bibitem{chen2017multi}
X.~Chen, H.~Ma, J.~Wan, B.~Li, and T.~Xia, ``{Multi-View 3D Object Detection
  Network for Autonomous Driving},'' in \emph{CVPR}, 2017.

\bibitem{xu2018pointfusion}
D.~Xu, D.~Anguelov, and A.~Jain, ``{PointFusion: Deep Sensor Fusion for 3D
  Bounding Box Estimation},'' in \emph{CVPR}, 2018, pp. 244--253.

\bibitem{liang2019multi}
M.~Liang, B.~Yang, Y.~Chen, R.~Hu, and R.~Urtasun, ``{Multi-Task Multi-Sensor
  Fusion for 3D Object Detection},'' in \emph{CVPR}, 2019.

\bibitem{du2017car}
X.~Du, M.~H. Ang, and D.~Rus, ``{Car Detection for Autonomous Vehicle: LIDAR
  and Vision Fusion Approach Through Deep Learning Framework},'' in
  \emph{IROS}, 2017.

\bibitem{ku2018joint}
J.~Ku, M.~Mozifian, J.~Lee, A.~Harakeh, and S.~L. Waslander, ``{Joint 3D
  Proposal Generation and Object Detection from View Aggregation},'' in
  \emph{IROS}, 2018, pp. 1--8.

\bibitem{ma2019accurate}
X.~Ma, Z.~Wang, H.~Li, P.~Zhang, W.~Ouyang, and X.~Fan, ``{Accurate Monocular
  3D Object Detection via Color-Embedded 3D Reconstruction for Autonomous
  Driving},'' in \emph{CVPR}, 2019, pp. 6851--6860.

\bibitem{du2018general}
X.~Du, M.~H. Ang, S.~Karaman, and D.~Rus, ``{A General Pipeline for 3D
  Detection of Vehicles},'' in \emph{ICRA 2018}.\hskip 1em plus 0.5em minus
  0.4em\relax IEEE, 2018, pp. 3194--3200.

\bibitem{quinonezsavior}
R.~Quinonez, J.~Giraldo, L.~Salazar, and E.~Bauman, ``{SAVIOR: Securing
  Autonomous Vehicles with Robust Physical Invariants},'' in \emph{USENIX
  Security}, 2020.

\bibitem{guo2018roboads}
P.~Guo, H.~Kim, N.~Virani, J.~Xu, M.~Zhu, and P.~Liu, ``{RoboADS: Anomaly
  Detection against Sensor and Actuator Misbehaviors in Mobile Robots},'' in
  \emph{DSN}.\hskip 1em plus 0.5em minus 0.4em\relax IEEE, 2018, pp. 574--585.

\bibitem{shin2017illusion}
H.~Shin, D.~Kim, Y.~Kwon, and Y.~Kim, ``{Illusion and Dazzle: Adversarial
  Optical Channel Exploits against LiDARs for Automotive Applications},'' in
  \emph{CHES}.\hskip 1em plus 0.5em minus 0.4em\relax Springer, 2017, pp.
  445--467.

\bibitem{engelcke2017vote3deep}
M.~Engelcke, D.~Rao, D.~Z. Wang, C.~H. Tong, and I.~Posner, ``{Vote3deep: Fast
  Object Detection in 3D Point Clouds using Efficient Convolutional Neural
  Networks},'' in \emph{ICRA}, 2017.

\bibitem{zhou2018voxelnet}
Y.~Zhou and O.~Tuzel, ``Voxelnet: End-to-end learning for point cloud based 3d
  object detection,'' in \emph{CVPR}, 2018, pp. 4490--4499.

\bibitem{lang2019pointpillars}
A.~H. Lang, S.~Vora, H.~Caesar, L.~Zhou, J.~Yang, and O.~Beijbom,
  ``{Pointpillars: Fast Encoders for Object Detection from Point Clouds},'' in
  \emph{CVPR}, 2019, pp. 12\,697--12\,705.

\bibitem{beltran2018birdnet}
J.~Beltran, C.~Guindel, F.~M. Moreno, D.~Cruzado, F.~Garcia, and
  A.~De~La~Escalera, ``{Birdnet: a 3D Object Detection Framework from Lidar
  Information},'' in \emph{ITSC}.\hskip 1em plus 0.5em minus 0.4em\relax IEEE,
  2018, pp. 3517--3523.

\bibitem{yang2018pixor}
B.~Yang, W.~Luo, and R.~Urtasun, ``{Pixor: Real-time 3d object detection from
  point clouds},'' in \emph{CVPR 2018}, 2018, pp. 7652--7660.

\bibitem{ranking-AD}
``{Navigant Research Names Waymo, Ford Autonomous Vehicles, Cruise, and Baidu
  the Leading Developers of Automated Driving Systems},''
  \url{https://www.businesswire.com/news/home/20200407005119/en/Navigant-Research-Names-Waymo-Ford-Autonomous-Vehicles}.

\bibitem{kittidataset}
A.~Geiger, P.~Lenz, C.~Stiller, and R.~Urtasun, ``{Vision Meets Robotics: The
  KiTTi Dataset},'' \emph{IJRR}, 2013.

\bibitem{ourwebsite}
``{Our Project Website},'' \url{https://sites.google.com/view/cav-sec/msf-adv}.

\bibitem{ivanov2014attack}
R.~Ivanov, M.~Pajic, and I.~Lee, ``{Attack-resilient Sensor Fusion},'' in
  \emph{DATE}.\hskip 1em plus 0.5em minus 0.4em\relax IEEE, 2014, pp. 1--6.

\bibitem{xu2018analyzing}
W.~Xu, C.~Yan, W.~Jia, X.~Ji, and J.~Liu, ``{Analyzing and Enhancing the
  Security of Ultrasonic Sensors for Autonomous Vehicles},'' \emph{IEEE
  Internet of Things Journal}, vol.~5, no.~6, pp. 5015--5029, 2018.

\bibitem{yan2016can}
C.~Yan, W.~Xu, and J.~Liu, ``{Can You Trust Autonomous Vehicles: Contactless
  Attacks against Sensors of Self-driving Vehicle},'' \emph{DEF CON}, vol.~24,
  no.~8, p. 109, 2016.

\bibitem{petit2014potential}
J.~Petit and S.~E. Shladover, ``{Potential Cyberattacks on Automated
  Vehicles},'' \emph{IEEE ITS}, vol.~16, no.~2, pp. 546--556, 2014.

\bibitem{man2020ghostimage}
Y.~Man, M.~Li, and R.~Gerdes, ``{GhostImage: Perception Domain Attacks against
  Vision-based Object Classification Systems},'' \emph{arXiv preprint
  arXiv:2001.07792}, 2020.

\bibitem{rearchitecting}
V.~Chandrasekaran, B.~Tang, N.~Papernot, K.~Fawaz, S.~Jha, and X.~Wu,
  ``{Rearchitecting Classification Frameworks For Increased Robustness},'' in
  \emph{arXiv:1905.10900}, 2019.

\bibitem{obj-detect-perf}
Z.-Q. Zhao, P.~Zheng, S.-t. Xu, and X.~Wu, ``{Object Detection with Deep
  Learning: A Review},'' \emph{IEEE NNLS}, pp. 3212--3232, 2019.

\bibitem{modular-end2end}
E.~Yurtsever, J.~Lambert, A.~Carballo, and K.~Takeda, ``{A Survey of Autonomous
  Driving: Common Practices and Emerging Technologies},'' \emph{IEEE Access},
  vol.~8, pp. 58\,443--58\,469, 2020.

\bibitem{chi1708deep}
L.~Chi and Y.~Mu, ``{Deep Steering: Learning End-to-End Driving Model from
  Spatial and Temporal Visual Cues},'' \emph{arXiv}, 2017.

\bibitem{pcsurvey}
Y.~Guo, H.~Wang, Q.~Hu, H.~Liu, L.~Liu, and M.~Bennamoun, ``{Deep Learning for
  3D Point Clouds: A Survey},'' 2019.

\bibitem{lidar-128}
``{Velodyne Alpha Prime},''
  \url{https://autonomoustuff.com/product/velodyne-vls-128/}.

\bibitem{wang2015voting}
D.~Z. Wang and I.~Posner, ``{Voting for Voting in Online Point Cloud Object
  Detection},'' in \emph{Robotics: Science and Systems}, 2015.

\bibitem{goodfellow:fsgm}
I.~Goodfellow, J.~Shlens, and C.~Szegedy, ``{Explaining and Harnessing
  Adversarial Examples},'' in \emph{ICLR}, 2015.

\bibitem{papernot:jsma}
N.~Papernot, P.~D. McDaniel, S.~Jha, M.~Fredrikson, Z.~B. Celik, and A.~Swami,
  ``{The Limitations of Deep Learning in Adversarial Settings},'' in \emph{Euro
  S\&P}, 2016.

\bibitem{xiao2018generating}
C.~Xiao, B.~Li, J.-Y. Zhu, W.~He, M.~Liu, and D.~Song, ``{Generating
  Adversarial Examples with Adversarial Networks},'' \emph{ArXiv}, 2018.

\bibitem{xiao2018spatially}
C.~Xiao, J.-Y. Zhu, B.~Li, W.~He, M.~Liu, and D.~Song, ``{Spatially Transformed
  Adversarial Examples},'' \emph{ICLR}, 2018.

\bibitem{xiao2018characterizing}
C.~Xiao, R.~Deng, B.~Li, F.~Yu, M.~Liu, and D.~Song, ``{Characterizing
  Adversarial Examples based on Spatial Consistency Information for Semantic
  Segmentation},'' in \emph{ECCV}, 2018, pp. 217--234.

\bibitem{xiao2019characterizing}
C.~Xiao, X.~Pan, W.~He, J.~Peng, M.~Sun, J.~Yi, M.~Liu, B.~Li, and D.~Song,
  ``{Characterizing Attacks on Deep Reinforcement Learning},'' \emph{arXiv},
  2019.

\bibitem{qiu2020semanticadv}
H.~Qiu, C.~Xiao, L.~Yang, X.~Yan, H.~Lee, and B.~Li, ``{SemanticAdv: Generating
  Adversarial Examples via Attribute-conditioned Image Editing},'' in
  \emph{ECCV}.\hskip 1em plus 0.5em minus 0.4em\relax Springer, 2020, pp.
  19--37.

\bibitem{li2019adversarial}
J.~Li, F.~Schmidt, and Z.~Kolter, ``{Adversarial Camera Stickers: A Physical
  Camera-based Attack on Deep Learning Systems},'' in \emph{ICML}, 2019, pp.
  3896--3904.

\bibitem{adv-patch}
T.~B. Brown, D.~Mané, A.~Roy, M.~Abadi, and J.~Gilmer, ``{Adversarial
  Patch},'' in \emph{arXiv:1712.09665}, 2017.

\bibitem{thys2019fooling}
S.~Thys, W.~Van~Ranst, and T.~Goedem{\'e}, ``{Fooling Automated Surveillance
  Cameras: Adversarial Patches to Attack Person Detection},'' in \emph{CVPR
  Workshops}, 2019, pp. 0--0.

\bibitem{athalye2017synthesizing}
A.~Athalye and I.~Sutskever, ``{Synthesizing Robust Adversarial Examples},'' in
  \emph{International Conference on Machine Learning (ICML)}, 2018.

\bibitem{aeb-fail}
``{Does your car have automated emergency braking? It's a big fail for
  pedestrians},''
  https://www.zdnet.com/article/does-your-car-have-automated-emergency-braking-its-a-big-fail-for-pedestrians/,
  2019.

\bibitem{eykholt2018robust}
K.~Eykholt, I.~Evtimov, E.~Fernandes, B.~Li, A.~Rahmati, C.~Xiao, A.~Prakash,
  T.~Kohno, and D.~Song, ``{Robust Physical-World Attacks on Deep Learning
  Visual Classification},'' in \emph{CVPR}, 2018.

\bibitem{avis-waymo}
``{Avis will Service Waymo’s Self-driving Minivans},''
  \url{https://www.theverge.com/2017/6/26/15873236/avis-waymo-google-self-driving-cars-vans}.

\bibitem{reverse-tesla}
``{Experimental Security Research of Tesla Autopilot},''
  \url{https://keenlab.tencent.com/en/whitepapers/Experimental_Security_Research_of_Tesla_Autopilot.pdf}.

\bibitem{madry:towards}
A.~Madry, A.~Makelov, L.~Schmidt, D.~Tsipras, and A.~Vladu, ``{Towards Deep
  Learning Models Resistant to Adversarial Attacks},'' in \emph{ICLR}, 2018.

\bibitem{carlini:cw}
N.~Carlini and D.~A. Wagner, ``{Towards Evaluating the Robustness of Neural
  Networks},'' in \emph{IEEE S\&P}, 2017.

\bibitem{3d-print}
``{3D Printing Online},'' \url{https://formlabs.com/software/}.

\bibitem{eykholt2019designing}
K.~Eykholt, ``{Designing and Evaluating Physical Adversarial Attacks and
  Defenses for Machine Learning Algorithms},'' Ph.D. dissertation, 2019.

\bibitem{huang2020spaa}
B.~Huang and H.~Ling, ``{SPAA: Stealthy Projector-based Adversarial Attacks on
  Deep Image Classifiers},'' \emph{ArXiv}, 2020.

\bibitem{traffic-cone-1m}
``{Traffic Cone},'' \url{https://en.wikipedia.org/wiki/Traffic_cone}.

\bibitem{ray-casting-intersection}
``{Intro to Rendering, Ray Casting},''
  \url{https://ocw.mit.edu/courses/electrical-engineering-and-computer-science/6-837-computer-graphics-fall-2012/lecture-notes/MIT6_837F12_Lec11.pdf}.

\bibitem{rendering}
H.~Kato, Y.~Ushiku, and T.~Harada, ``{Neural 3D Mesh Renderer},'' in
  \emph{CVPR}, June 2018.

\bibitem{maturana2015voxnet}
D.~Maturana and S.~Scherer, ``Voxnet: A 3d convolutional neural network for
  real-time object recognition,'' in \emph{IROS}, 2015, pp. 922--928.

\bibitem{trilinear}
``{Tri. Interpolation},'' \url{en.wikipedia.org/wiki/Trilinear_interpolation}.

\bibitem{laplacian1}
S.~Li, X.~Xu, L.~Nie, and T.-S. Chua, ``{Laplacian-Steered Neural Style
  Transfer},'' in \emph{ICMR}, 2017, pp. 1716--1724.

\bibitem{brake-distance}
``{Brake Distance},'' \url{http://www.csgnetwork.com/stopdistcalc.html}.

\bibitem{carma-platform}
``{Carma Platform},'' \url{https://github.com/usdot-fhwa-stol/carma-platform}.

\bibitem{autoware-car}
``{Autoware Self-driving Vehicle on a Highway},''
  \url{https://www.youtube.com/watch?v=npQMzH3j_d8}.

\bibitem{Apollo-lincon}
``{Baidu launches their open platform for autonomous cars–and we got to test
  it},''
  \url{https://technode.com/2017/07/05/baidu-apollo-1-0-autonomous-cars-we-test-it/}.

\bibitem{apollo-taxi}
``{Baidu Launches Public Robotaxi Trial Operation},''
  \url{https://www.globenewswire.com/news-release/2019/09/26/1921380/0/en/Baidu-Launches-Public-Robotaxi-Trial-Operation.html}.

\bibitem{yolo-darknet}
``{YOLOv3 Darknet},'' \url{https://pjreddie.com/darknet/yolo/}.

\bibitem{zhang2018perceptual}
R.~Zhang, P.~Isola, A.~A. Efros, E.~Shechtman, and O.~Wang, ``{The Unreasonable
  Effectiveness of Deep Features as a Perceptual Metric},'' in \emph{CVPR},
  2018.

\bibitem{jo2020investigating}
Y.~Jo, S.~Yang, and S.~Joo~Kim, ``{Investigating Loss Functions for Extreme
  Super-Resolution},'' in \emph{CVPR Workshops}, 2020, pp. 424--425.

\bibitem{MTurk}
``{Amazon Mechanical Turk},'' \url{https://www.mturk.com}.

\bibitem{aaai3dobject}
T.~Tsai, K.~Yang, T.-Y. Ho, and Y.~Jin, ``{Robust Adversarial Objects against
  Deep Learning Models},'' in \emph{AAAI}, 2020.

\bibitem{formlabs}
``{FormLabs},'' \url{https://formlabs.com/software/}.

\bibitem{gaussian-curvature}
``{Curvature},'' \url{https://en.wikipedia.org/wiki/Gaussian_curvature}.

\bibitem{mitchell1998introduction}
M.~Mitchell, \emph{An introduction to genetic algorithms}.\hskip 1em plus 0.5em
  minus 0.4em\relax MIT press, 1998.

\bibitem{gaussiannoise}
``{Adding Gaussian Noise},''
  \url{https://pytorch.org/docs/stable/tensors.html}.

\bibitem{geneticalgorithm}
``{Genetic Algorithm},'' \url{https://pypi.org/project/geneticalgorithm/}.

\bibitem{alzantot2019genattack}
M.~Alzantot, Y.~Sharma, S.~Chakraborty, H.~Zhang, C.-J. Hsieh, and M.~B.
  Srivastava, ``{Genattack: Practical Black-box Attacks with Gradient-free
  Optimization},'' in \emph{GECCO}, 2019.

\bibitem{Feng2020CGATTACKMT}
Y.~Feng, B.~Wu, Y.~Fan, L.~Liu, Z.~Li, and S.~Xia, ``{CG-ATTACK: Modeling the
  Conditional Distribution of Adversarial Perturbations to Boost Black-Box
  Attack},'' 2020.

\bibitem{alika2020optimization-10cm-1}
R.~Alika, E.~M. Mellouli, and E.~H. Tissir, ``{Optimization of Higher-Order
  Sliding Mode Control Parameter using Particle Swarm Optimization for Lateral
  Dynamics of Autonomous Vehicles},'' in \emph{IRASET}.\hskip 1em plus 0.5em
  minus 0.4em\relax IEEE, 2020, pp. 1--6.

\bibitem{dominguez2016comparison-10cm-2}
S.~Dominguez, A.~Ali, G.~Garcia, and P.~Martinet, ``{Comparison of Lateral
  Controllers for Autonomous Vehicle: Experimental Results},'' in \emph{2016
  IEEE 19th International Conference on Intelligent Transportation Systems
  (ITSC)}.\hskip 1em plus 0.5em minus 0.4em\relax IEEE, 2016, pp. 1418--1423.

\bibitem{large-format-3d-print}
``{LARGE-FORMAT 3D PRINTER FOR INDUSTRIAL APPLICATIONS},''
  \url{https://bigrep.com/bigrep-one/}.

\bibitem{lanewidth}
\BIBentryALTinterwordspacing
AASHTO, \emph{Policy on Geometric Design of Highways and Streets (7th
  Edition)}, 2018. [Online]. Available:
  \url{https://app.knovel.com/hotlink/toc/id:kpPGDHSE12/policy-geometric-design/policy-geometric-design}
\BIBentrySTDinterwordspacing

\bibitem{apollo-github}
``{Apollo Models},''
  \url{https://github.com/ApolloAuto/apollo/tree/r5.5.0/modules/perception/production/data/perception/lidar/models/cnnseg}.

\bibitem{LGSVL}
``{LGSVL Simulator},'' \url{https://www.lgsvlsimulator.com/}.

\bibitem{company-simulation2}
``{Inside Waymo's Secret World for Training Self-Driving Cars},''
  \url{https://www.theatlantic.com/technology/archive/2017/08/inside-waymos-secret-testing-and-simulation-facilities/537648/}.

\bibitem{Xu:2018:ndss}
W.~{Xu}, D.~{Evans}, and Y.~{Qi}, ``{Feature Squeezing: Detecting Adversarial
  Examples in Deep Neural Networks},'' in \emph{NDSS}, 2018.

\bibitem{meng2017magnet}
D.~Meng and H.~Chen, ``{MagNet: a Two-Pronged Defense against Adversarial
  Examples},'' in \emph{ACM CCS}, 2017, pp. 135--147.

\bibitem{dziugaite2016study}
G.~K. Dziugaite, Z.~Ghahramani, and D.~M. Roy, ``{A Study of the Effect of JPG
  Compression on Adversarial Images},'' \emph{arXiv}, 2016.

\bibitem{li2020sok}
L.~Li, X.~Qi, T.~Xie, and B.~Li, ``{SoK: Certified Robustness for Deep Neural
  Networks},'' \emph{arXiv preprint arXiv:2009.04131}, 2020.

\bibitem{lecuyer2019certified}
M.~Lecuyer, V.~Atlidakis, R.~Geambasu, D.~Hsu, and S.~Jana, ``{Certified
  Robustness to Adversarial Examples with Differential Privacy},'' in
  \emph{IEEE S\&P}.

\bibitem{szegedy:iclr:2014}
C.~Szegedy, W.~Zaremba, I.~Sutskever, J.~Bruna, D.~Erhan, I.~Goodfellow, and
  R.~Fergus, ``{Intriguing Properties of Neural Networks},'' in \emph{ICLR},
  2014.

\bibitem{zhang2019towards}
H.~Zhang and J.~Wang, ``{Towards Adversarially Robust Object Detection},'' in
  \emph{ICCV}, 2019, pp. 421--430.

\bibitem{wong2020fast}
E.~Wong, L.~Rice, and J.~Z. Kolter, ``{Fast is Better than Free: Revisiting
  Adversarial Training},'' \emph{ICLR}, 2020.

\bibitem{cohen2019certified}
J.~M. Cohen, E.~Rosenfeld, and J.~Z. Kolter, ``{Certified Adversarial
  Robustness via Randomized Smoothing},'' \emph{ICML}, 2019.

\bibitem{yang2020randomized}
G.~Yang, T.~Duan, E.~Hu, H.~Salman, I.~Razenshteyn, and J.~Li, ``{Randomized
  Smoothing of All Shapes and Sizes},'' \emph{ICML}, 2020.

\bibitem{carlini2017magnet}
N.~Carlini and D.~Wagner, ``{MagNet and "Efficient Defenses against Adversarial
  Attacks" are not Robust to Adversarial Examples},'' \emph{arXiv}, 2017.

\bibitem{zhang2020interpretable}
X.~Zhang, N.~Wang, H.~Shen, S.~Ji, X.~Luo, and T.~Wang, ``{Interpretable Deep
  Learning under Fire},'' in \emph{USENIX Security}, 2020.

\bibitem{he2017adversarial}
W.~He, J.~Wei, X.~Chen, N.~Carlini, and D.~Song, ``{Adversarial Example
  Defense: Ensembles of Weak Defenses are not Strong},'' in \emph{USENIX WOOT},
  2017.

\bibitem{sharma2018bypassing}
Y.~Sharma and P.-Y. Chen, ``{Bypassing Feature Squeezing by Increasing
  Adversary Strength},'' \emph{ICLR Workshop}, 2018.

\bibitem{wang2019study}
L.~Wang, J.~Tang, and Q.~Liao, ``{A Study on Radar Target Detection based on
  Deep Neural Networks},'' \emph{IEEE Sensors Letters}, pp. 1--4, 2019.

\bibitem{nassi2020phantom}
B.~Nassi, D.~Nassi, R.~Ben-Netanel, Y.~Mirsky, O.~Drokin, and Y.~Elovici,
  ``{Phantom of the ADAS: Phantom Attacks on Driver-Assistance Systems},'' in
  \emph{IACR}, 2020.

\bibitem{tu2018injected}
Y.~Tu, Z.~Lin, I.~Lee, and X.~Hei, ``{Injected and Delivered: Fabricating
  Implicit Control over Actuation Systems by Spoofing Inertial Sensors},'' in
  \emph{USENIX Security}, 2018, pp. 1545--1562.

\bibitem{povolny2020adas}
``{Model Hacking ADAS to Pave Safer Roads for Autonomous Vehicles},''
  \url{https://www.mcafee.com/blogs/other-blogs/mcafee-labs/model-hacking-adas-to-pave-safer-roads-for-autonomous-vehicles/},
  2020.

\bibitem{jia2019fooling}
Y.~Jia, Y.~Lu, J.~Shen, Q.~A. Chen, H.~Chen, Z.~Zhong, and T.~Wei, ``{Fooling
  Detection Alone is Not Enough: Adversarial Attack Against Multiple Object
  Tracking},'' in \emph{ICLR}, 2019.

\bibitem{junjie:usenix:2020}
J.~Shen, J.~Y. Won, and Q.~A. Chen, ``{{Drift with Devil: Security of
  Multi-Sensor Fusion based Localization in High-Level Autonomous Driving under
  GPS Spoofing}},'' in \emph{Usenix Security}, 2020.

\bibitem{sato2020hold}
T.~Sato, J.~Shen, N.~Wang, Y.~J. Jia, X.~Lin, and Q.~A. Chen, ``{Hold Tight and
  Never Let Go: Security of Deep Learning based Automated Lane Centering under
  Physical-World Attack},'' \emph{ArXiv}, 2020.

\bibitem{sato2021wip}
T.~Sato, J.~Shen, N.~Wang, Y.~J. Jia, X.~Lin, and Q.~A. Chen, ``{Deployability
  Improvement, Stealthiness User Study, and Safety Impact Assessment on Real
  Vehicle for Dirty Road Patch Attack},'' in \emph{AutoSec Workshop at NDSS},
  2021.

\bibitem{liang2021wip}
H.~Liang, R.~Jiao, T.~Sato, J.~Shen, Q.~A. Chen, and Q.~Zhu, ``{End-to-End
  Analysis of Adversarial Attacks to Automated Lane Centering Systems},'' in
  \emph{AutoSec Workshop at NDSS}, 2021.

\bibitem{kanglan2021}
K.~Tang, J.~Shen, and Q.~A. Chen, ``{Fooling Perception via Location: A Case of
  Region-of-Interest Attacks on Traffic Light Detection in Autonomous
  Driving},'' in \emph{AutoSec Workshop at NDSS}, 2021.

\bibitem{moosavi2016deepfool}
S.-M. Moosavi-Dezfooli, A.~Fawzi, and P.~Frossard, ``{DeepFool: a Simple and
  Accurate Method to Fool Deep Neural Networks},'' in \emph{CVPR}, 2016, pp.
  2574--2582.

\bibitem{xiao2019meshadv}
C.~Xiao, D.~Yang, B.~Li, J.~Deng, and M.~Liu, ``{MeshAdv: Adversarial Meshes
  for Visual Recognition},'' in \emph{IEEE CVPR}, 2019.

\bibitem{xiang2019generating}
C.~Xiang, C.~R. Qi, and B.~Li, ``{Generating 3D Adversarial Point Clouds},'' in
  \emph{CVPR}, 2019, pp. 9136--9144.

\bibitem{lee2020shapeadv}
K.~Lee, Z.~Chen, X.~Yan, R.~Urtasun, and E.~Yumer, ``{ShapeAdv: Generating
  Shape-Aware Adversarial 3D Point Clouds},'' \emph{ArXiv}, 2020.

\bibitem{printer-precision}
``{Understanding Accuracy, Precision, and Tolerance in 3D Printing},''
  \url{https://formlabs.com/blog/understanding-accuracy-precision-tolerance-in-3d-printing/}.

\bibitem{userstudydrive}
``{User Study: Anomalous Traffic Cone Survey},''
  \url{https://drive.google.com/file/d/1EqtQL6m1ZPNOQGs6pbAM25WFT8D58EC2/view}.

\bibitem{QECD}
``{Mesh Simplification},''
  \url{http://graphics.stanford.edu/courses/cs468-10-fall/LectureSlides/08_Simplification.pdf}.

\bibitem{pillow}
``{Pillow (PIL Fork)},'' \url{https://pillow.readthedocs.io/en/stable/}.

\bibitem{shafahi2019adversarial}
A.~Shafahi, M.~Najibi, M.~A. Ghiasi, Z.~Xu, J.~Dickerson, C.~Studer, L.~S.
  Davis, G.~Taylor, and T.~Goldstein, ``{Adversarial Training for Free!}'' in
  \emph{NIPS}, 2019, pp. 3358--3369.

\bibitem{hendrycks2019using}
D.~Hendrycks, K.~Lee, and M.~Mazeika, ``{Using Pre-Training can Improve Model
  Robustness and Uncertainty},'' \emph{ICML}, 2019.

\bibitem{chen2020adversarial}
T.~Chen, S.~Liu, S.~Chang, Y.~Cheng, L.~Amini, and Z.~Wang, ``{Adversarial
  Robustness: From Self-Supervised Pre-Training to Fine-Tuning},'' in
  \emph{CVPR}, 2020, pp. 699--708.

\bibitem{coco}
``{COCO Dataset},'' \url{http://cocodataset.org/}.

\end{thebibliography}

\appendix
\label{sec:appendix}

\begin{table}[htb]
\centering
			\begin{tabular}{lc}
\toprule
			Parameter  &  Value\\
				\midrule
			PGD initial point (\msec{sec:objective-function})&0.01  \\
			PGD constraint (\msec{sec:objective-function}) &0.02  \\
			Tanh approximation parameter $\mu$ (\msec{sec:preprocessing})& 100\\
			Preventing division by zero $\varepsilon$ (\msec{sec:preprocessing})& $10^{-7}$\\
			$X$ sample range (\msec{sec:objective-function})& $(5, 35)$\\
			$Y$ sample range (\msec{sec:objective-function})& $(-0.3, 0.3)$\\
			$yaw$ sample angles (\msec{sec:objective-function})& $(-5^{\circ}, 5^{\circ}) $\\
			Learning rate (\msec{sec:objective-function}) & $ 0.001$\\
			$\gL_{r}(\cdot)$ coefficient $\lambda$ (\msec{sec:objective-function}) & $20$\\
			Height loss coefficient $\beta_1$ (Appendix \ref{sec:realize-loss}) & $ 0.001$\\
			Precision of 3D printer used in~\S\ref{sec:attack-realizability}  & 0.38mm\\
				\bottomrule
			\end{tabular}
     \caption{Detailed settings for attack parameters in \S\ref{sec:evaluation}. }
    \label{tab:para-setting}
\vspace{-0.4cm}
\end{table}

\subsection{Realizability loss $\gL_r(\cdot)$ in~\S\ref{sec:objective-function}}
\label{sec:realize-loss}

To realize our attack goal in~\S\ref{sec:attack-goal}, $S^a$ needs to be 3D-printed and placed on top of the road surface in the physical world. To facilitate this, we design the realizability loss $\gL_r(\cdot)$ in our objective function to (1) improve the printability of $S^a$ by 3D printers, and (2) prevent the generation of $S^a$ that is underneath the road surface. Our formulation of $\gL_r(\cdot)$ is in \meq{eq:pecp-loss}, where the first and second parts are for achieving (1) and (2) respectively. The first part is a Laplacian loss\mcite{laplacian1}, where $V^a$ is the vertex set of $S^a$, and for $\vv^a_i \in V^a$, $\Gamma(\vv^a_i)$ denotes the set of connected neighboring vertices of $\vv^a_i$. Since our attack generation is performed by only moving the vertex positions in the benign object $S$ (\S\ref{sec:design-overview}), there is always a corresponding vertex $\vv_i$ in the vertex set $V$ of $S$ that $\vv^a_i$ is moved from. The distance between $\vv^a_i$ and $\vv_i$ is denoted as $\Delta\vv=\vv^a_i - \vv_i$. Thus, the first part in \meq{eq:pecp-loss} penalties the differences between the position change of each vertex in $S^a$ and those of its neighboring vertices. This can thus improve the smoothness of the surface of $S^a$, which can lower the precision requirements of the 3D printer\mcite{printer-precision} and thus improve the printability of $S^a$. We also use a popular mesh simplification method, Quadric Edge Collapse Decimation (QECD), as an optional post-processing step to further improve printability. 

In the second part, $z^a_i$ and $z_i$ denotes the height values of $\vv^a_i$ and $\vv_i$. This part minimizes the distance between the lowest height among all vertices in $S^a$ and that in $S$, which thus penalties the moving of the vertices in $S^a$ towards under the road surface. $\beta_1$ is a hyper-parameter for this part in \meq{eq:pecp-loss}.

\begin{footnotesize}
\begin{equation}
\vspace{-0.5cm}
\begin{split}
    \label{eq:pecp-loss}
    \gL_{r} (S^a, S)  = & \sum_{ \vv^a_i \in V^a} \sum_{\vv^a_q \in{\Gamma(\vv^a_i)}} \norm{\Delta\vv^a_i - \Delta\vv^a_q}^2_2  \\& + \beta_1 \cdot \|\min_{\vv^a_i \in V^a}{z^a_i} - \min_{\vv_i \in V}{z_i}\|^2_2
\end{split} 
\end{equation}  
\end{footnotesize}

\subsection{Attack Stealthiness User Study}
\label{sec:stealthiness-user-study}

In this section, we conduct a user study to evaluate the stealthiness of the adversarial 3D objects. We go through the IRB process and our study is determined as the IRB Exempt, due to not involving collection of any Personally Identifiable Information (PII) or target any sensitive population.

\textbf{Evaluation methodology.} In this study, we select traffic cone as the evaluation target due to its high attractiveness for the attacker (\S\ref{sec:eval_setup}). 
We evaluate 4 red traffic cone with different shapes: the benign shape (\textit{Benign}) the adversarial shape generated by \system (\textit{Adv}), and two benign but broken shapes similar to Fig.~\ref{fig:attack-vectors} (\textit{Benign B1} and \textit{B2}). We consider \textit{Benign B1} and \textit{B2} since our attack is designed to mimic benign traffic objects with a broken look (\S\ref{sec:design-overview}). We randomly select two images (\textit{S1}, \textit{S2}) from KITTI and render these shapes into these two images at two different positions (near or far way from the victim AV, denoted as \textit{N} or \textit{F}) to generate four realistic driver's scenario  (\textit{S1-N}, \textit{S1-F}, \textit{S2-N}, \textit{S2-F}) on the roadway for each shape. 

For each of the 4 rendered images above, we ask whether the red traffic object in the image is a valid traffic cone. Note that the driving images are selected by ensuring no extra red object. Images of them are on our website~\cite{ourwebsite}.
Among the 4 rendered images, we also ask in which one the red traffic object has the most anomalous shape compared to a normal traffic cone. ``No Anomaly'' option is included to avoid randomly picking from the participants.  
To understand the distribution of the participant's background, we also ask for demographic information and background information related to driving. 

\textbf{Evaluation setup.}
We use Amazon Mechanical Turk~\cite{MTurk} to perform the user study. In total, we collected results from 105 participants (55.24\%  male and 44.76\% female) with  35.3 average age.
We confirmed that all of them have driving experience by asking them the age when first licensed and the weekly driving mileage. 
All the benign objects, including \textit{Benign B1} and \textit{B2} can be correctly detected by the latest Apollo MSF combination (A5-L\textcircled{+}A5-C) while \textit{Adv} cannot. The full survey is available at~\cite{userstudydrive}.

\begin{figure}[t!]
    \begin{subfigure}{0.48\linewidth}
     \centering
     \includegraphics[width=1.\linewidth]{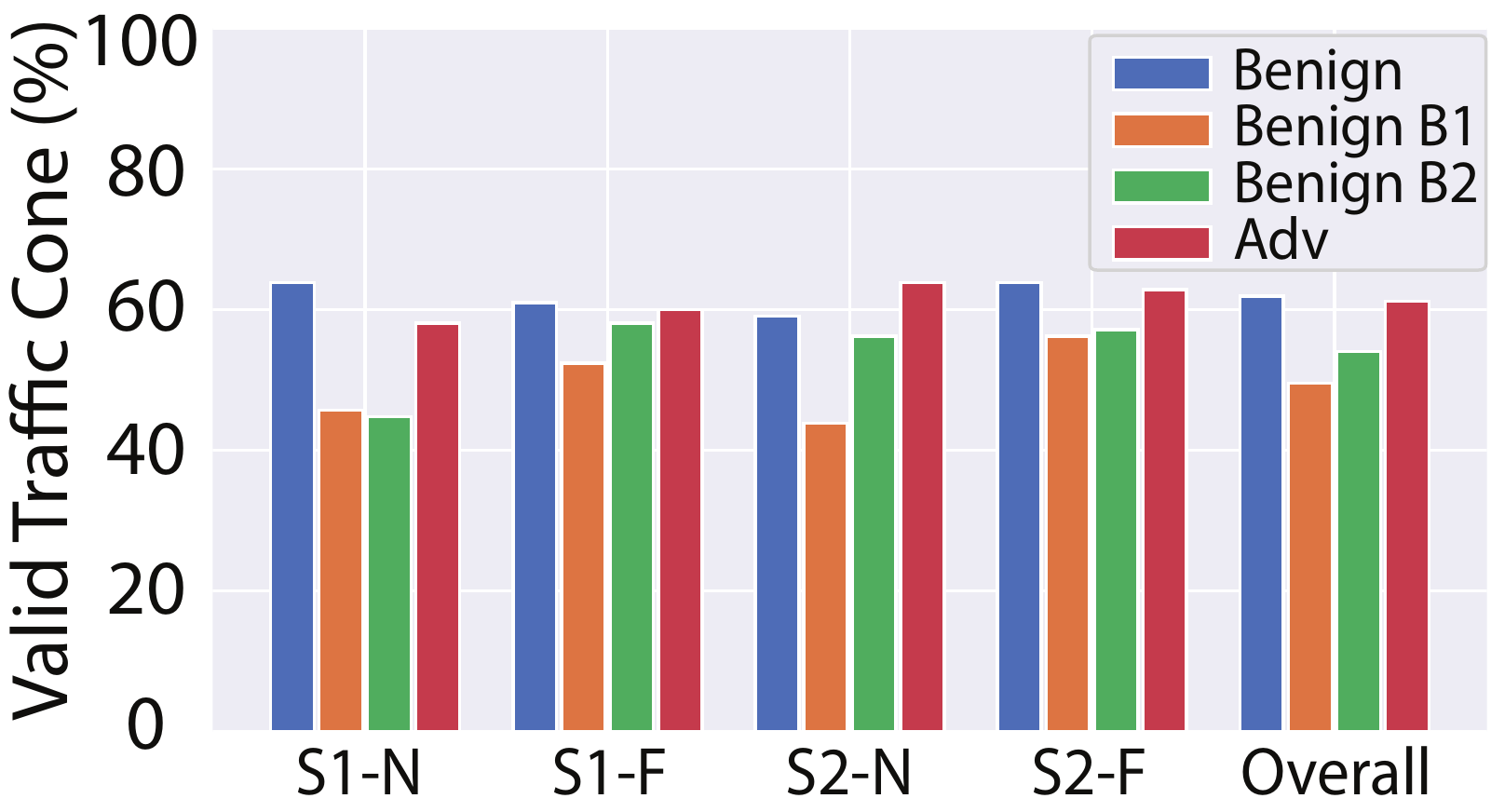}
     \caption{Validity of the traffic cone}
     \label{fig:USvalidCone}
     \end{subfigure}
     \begin{subfigure}{0.48\linewidth}
     \centering
        
     \includegraphics[width=\linewidth]{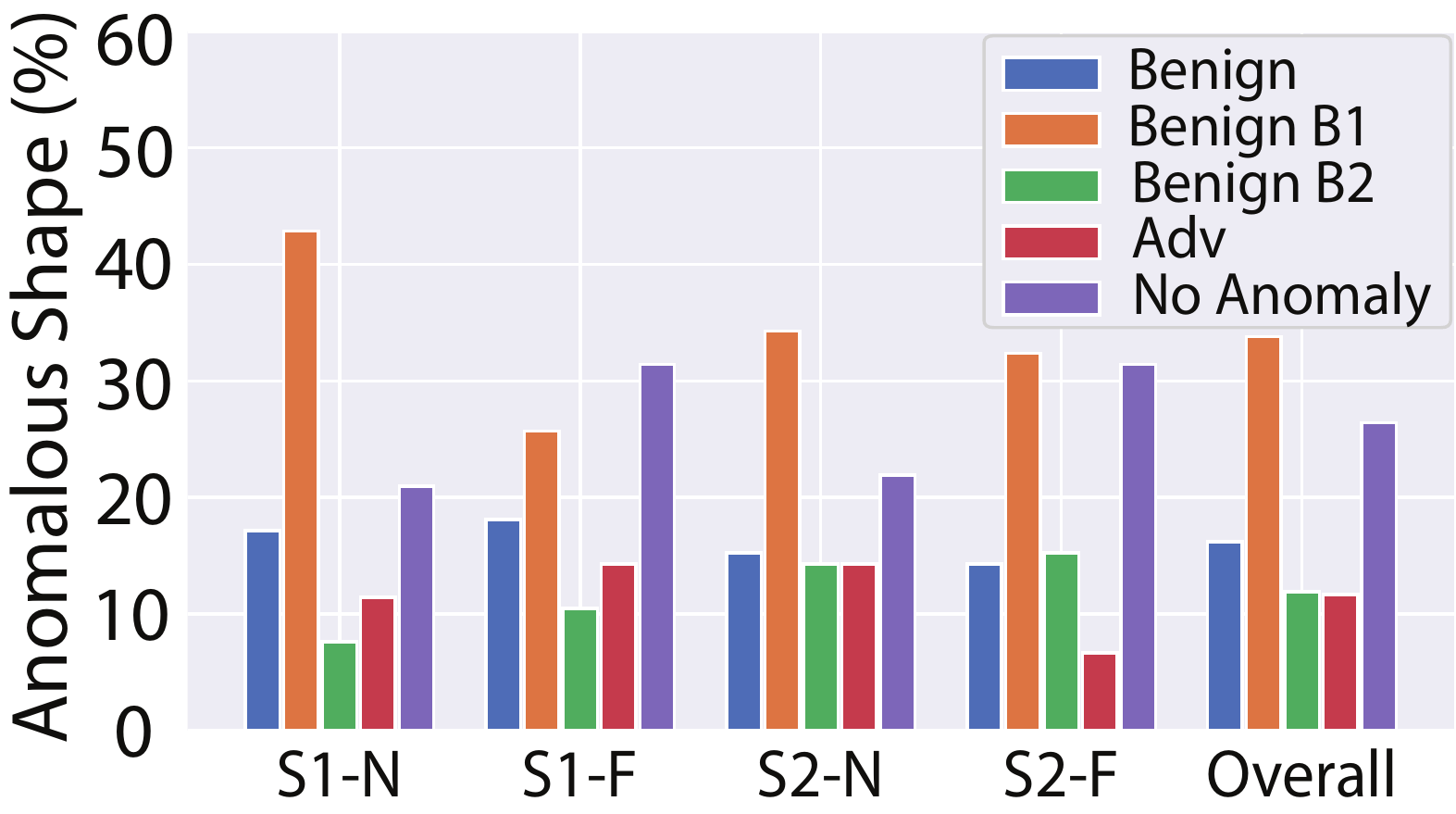}
     \caption{Most anomalous shape}
     \label{fig:USamomalyShape} 
     \end{subfigure}
     \caption{User study results of the attack stealthiness. (a) shows the ratio of users thinking a given object is a valid traffic cone; (b) shows the selection ratios for traffic cones with the most anomalous shape. \textit{S1} and \textit{S2}: 2 different real-world driving images; \textit{N} and \textit{F}: near and far rendering positions.}
     \label{fig:USres}
\end{figure}

\textbf{Results.} 
Fig.~\ref{fig:USvalidCone} shows the ratio of users thinking that the given traffic cone object is a valid traffic cone. As shown,
\textit{Benign} and \textit{Adv} have similar ratios (around 60\%) and are higher than \textit{Benign B1} and \textit{B2} since the broken shapes may be more obvious than that of \textit{Adv} after our surface smoothing and PGD-based perturbation bounding (\S\ref{sec:objective-function}). 
Note that even for \textit{Benign} there are around 40\% users thinking that it is invalid. This might be because the rendered color and shading inevitably have infidelity compared to the real-world background images. 
Fig.~\ref{fig:USamomalyShape} shows the selection ratios for the cone object with the most anomalous shape. As shown, \textit{Benign B1} and ``No Anomaly'' are the most popular choices across and \textit{Adv} is always the lowest. 
The results show that our generated adversarial traffic cone is generally viewed at least as innocent as the original benign cone, and also less suspicious than certain benign ones with broken shapes. 

\nsubsection{Attack Effectiveness under Different Attack Settings}
\label{sec:different-settings}

In this section, we perform experiments to understand how sensitive our attack is to different attack parameter settings. 

{\bf Experimental setup.} We target 5 key attack parameters in Table~\ref{tab:para-setting} to perform experiments: $\mu$,  $\lambda$, $\beta_1$, Learning rate, and PGD initial. 
For each parameter, we experiment with values that are one magnitude higher or lower than the default value. The experiments are performed on A5-L\textcircled{+}A5-C with traffic cone. The results are averaged over 20 attack scenarios randomly selected from the 100 scenarios in~\S\ref{sec:eval_setup}.

{\bf Results.}
Table~\ref{tab:different-para} shows the results. As shown, our attack is most sensitive to $\mu$. Considering that $\mu$ is used in our differentiable approximation of the point-inclusion calculation (\meq{eq:tanh} in~\S\ref{sec:preprocessing}), these results show that our differentiable approximation design is critical to the attack success. Different learning rates, $\lambda$, and PGD initial are also shown to impact the results, but such impacts are limited to when the values are above or below a certain magnitude.

\begin{table*}[t]
    \centering

\setlength\tabcolsep{4.0pt}
  \begin{tabular}{lccccccccccccccc}
    \toprule
    \centering
    & \multicolumn{3}{c}{$\mu$}  &  \multicolumn{3}{c}{$\lambda$} & \multicolumn{3}{c}{$\beta_1$} & \multicolumn{3}{c}{Learning rate} & \multicolumn{3}{c}{Initialization}
    \\
    \cmidrule(lr){2-4} \cmidrule(lr){5-7}
    \cmidrule(lr){8-10}
    \cmidrule(lr){11-13}
    \cmidrule(lr){14-16}
    
    \multirow{-1}{*}{Key attack parameters} & 10 & \cellcolor{Gray} 100 & 1000  & 2.0 &\cellcolor{Gray} 20.0 & 200.0 & 0.01 &\cellcolor{Gray} 0.001 & 0.0001 & 0.01 & \cellcolor{Gray} 0.001 & 0.0001 & 0.1 & \cellcolor{Gray} 0.01 & 0.001\\
    \midrule
        
       Attack success rate & 75\% & \cellcolor{Gray} 100\% & 60\%  &100\% & \cellcolor{Gray} 100\% & 65\% & 100\% & \cellcolor{Gray} 100\% & 100\% & 75\% & \cellcolor{Gray}100.0\% & 100\% & 45\% & \cellcolor{Gray}100.0\% & 100\%\\
        \bottomrule
    \end{tabular}
    \caption{Attack success rate for A5-L\textcircled{+}A5-C with traffic cone under different attack parameter settings. Descriptions of these attacker parameters are in Table~\ref{tab:para-setting}. Gray cells are default settings.}
    \label{tab:different-para}

\end{table*}

\nsubsection{Printability Evaluation}
\label{sec:attack-printability}
To perform our attack, the adversarial objects generated digitally need to be (1) printable by today's 3D printers, and (2) the easier to be printed the better, e.g., requiring less printing precision and thus printable by cheaper 3D printers. In this section, we evaluate such printability of our attack.

{\bf Evaluation metrics.}
To evaluate whether it is \textit{printable or not}, we first use \textit{PreForm}, a commercial printability checking tool~\cite{formlabs} that can determine whether their 3D-printing service can print a given 3D mesh. We also leverage the object \textit{watertightness}\mcite{aaai3dobject} as another metric, which measures whether the object mesh could hold water if filled. Thus, any 3D object needs to be watertight to have a volume and thus can validly exist (and thus 3D-printed) in physical world. This is the most basic metric for any object meshes to be printable.

For whether the object is \textit{easy to print}, we use the \textit{self-intersection ratio} and \textit{curvature}.  \textit{Self-intersection ratio} measures the percentage of the object mesh's 2D faces that have intersections with its other faces.  High self-intersection ratio means the mesh need to be printed by a higher precision printer with higher cost. 
The second metric we use is the \textit{curvature} of the object, which measures how smooth the object surface is. The more smooth the surface is, the less printing precision is required and thus the easier to print. We calculate this metric using the average per-vertex Gaussian curvature value.

{\bf Experimental setup.} As described in~\S\ref{sec:objective-function} and Appendix~\ref{sec:realize-loss}, our design includes two methods to improve the printability: the Laplacian loss (LP) in $\gL_r(\cdot)$ in \meq{eq:pecp-loss}, and QECD~\cite{QECD} as an optional post-processing step. Thus, in our experiment we evaluate the printability of our adversarial objects with and without these two methods. We use the same scenario and parameter settings as~\S\ref{sec:effectiveness-attack}.

\begin{table}[t]{\footnotesize
    \centering
    \setlength\tabcolsep{2.0pt}
  \begin{tabular}{cccccc}
    \toprule
        & \multicolumn{2}{c}{Printable?
        } & \multicolumn{2}{c}{Easy to Print?
        } &  \\ 
        \cmidrule(lr){2-3}
        \cmidrule(lr){4-5}
        \multirow{-2}{*}{\shortstack{Technique \\ used}} & {PreForm} & {Watertight} &{Self-intersect} & {Curvature} & \multirow{-2}{*}{\shortstack{Success \\ Rate}} \\
        \midrule
        None& $100\%$& $100\%$ & $88.73\%$& $1.68 \pm 1.56$ & $100\%$\\
        LP & $100\%$& $100\%$ & $14.43\%$& $0.69 \pm 0.65$ &$100\%$ \\
        QECD & 100\% & $100\%$&$38.96 \% $&$1.42 \pm 1.30$ & $90\%$\\
         LP + QECD  & $100\% $& $100\%$ & $0.46 \%$& $0.67 \pm 0.50$& $92\%$\\
        \bottomrule
    \end{tabular}
    \caption{Printability evaluation results of \system on A5-L\textcircled{+}A5-C with traffic cone. LP: Laplacian loss in \meq{eq:pecp-loss}.}
        \label{tab:Printability}
    }
\end{table}

{\bf Results.} Table\mref{tab:Printability} shows the evaluation results for A5-L\textcircled{+}A5-C using traffic cones. As shown, with or without using any printability improvement methods, the objects generated by our method are all watertight and determined as printable by the PreForm software since our attack method only manipulates the vertex positions of the benign object without changing the original vertex connection relationships. 

For the two metrics on whether the object is easy to print, both the self-intersection ratio and the average curvature value are greatly reduced by applying either LP or QECD. LP alone is particularly cost-effective: it is able to substantially reduce the self-intersection ratio by 74.3\% and the curvature value by 58.9\% without hurting the attack success rate.
However, QECD alone hurts self-intersect ratio, curvature value and attack success rate. 
The decrease of the attack success rate is because QECD is a mesh simplification method that may slightly change the object shape, which thus may interfere with the originally well-optimized shape of the adversarial object.  Combining QECD and LP together achieves the highest reduction in both metrics, with only 0.46\% self-intersection ratio and 0.67 curvature value. Note that, the average curvature for the benign traffic cone object  is 0.72. Thus, both LP alone and the combination achieve a similar level of surface smoothness comparable to a normal real-world object, which thus are printable enough in practice. While the combination  reduces the self-intersection ratio compared to LP alone, it incurs 8\% success rate decrease due to the use of QECD. Thus, there exists a trade-off. If the attacker does not care about the printing cost, they can choose to use LP alone to better ensure the attack success; otherwise, they can combine it with QECD to reduce the printing costs.

\subsection{Details of the DNN-Level Defenses Evaluated in~\S\ref{sec:defense-dnn}}
\label{sec:defense-setup}
\noindent We describe the details of the defense methods.

\noindent {\bf Bit-Depth reduction~\cite{Xu:2018:ndss}.} 
We follow the setting in prior work~\cite{Xu:2018:ndss}. We reduce the bit depth for the image input and the LiDAR point cloud. 
For a camera image, it consists of RGB channel with 8-bit depth (0-255) for each of them.
For a LiDAR point cloud, each point has 4 fields: $x$, $y$, $z$, $i$, where $x$, $y$, and $z$ represents the 3D position, and $i$ is intensity. Each field is a floating point with 32-bit. We use the formulation: $ \frac{ \mathrm{round}(x * (2^\text{bit} - 1))}{ (2^\text{bit}- 1)} $
to reduce the bit-depth.  
In our experiments, we evaluate 5 different bit-depths ranging from 5-bits to 1-bit for both camera and LiDAR inputs. Higher bit-depth number means higher input quality after the bit-depth reduction. 

\noindent {\bf Median smoothing~\cite{Xu:2018:ndss}.} 
We follow the setting in prior work~\cite{Xu:2018:ndss} and apply the median smoothing to both LiDAR and camera inputs by taking a median around each LiDAR point or camera pixel with a different kernel size. 
We evaluate 7 different kernel sizes ranging from 5 to 35.
Larger the kernel size means higher smoothness and lower input quality.

\noindent {\bf JPEG compression~\cite{dziugaite2016study}.}
We follow the setting in~\cite{dziugaite2016study} and apply the JPEG only to images since  JPEG compression is specific to images. Our attack is successful only when it succeeds for both camera and LiDAR models, therefore,  securing the camera model alone is still an effective defense strategy.
We use Python Image Library (PIL)~\cite{pillow} to control the compression quality with argument ``quality''. We use 9 values from 10 to 90 with step 10 to explore the defense effectiveness at different compression rates. Lower values means higher compression rates and thus lower image quality.

\noindent {\bf Autoencoder reformation~\cite{meng2017magnet}.}
Autoencoder reformation is a part of MagNet defense~\cite{meng2017magnet}. We apply it only on image since it is designed for camera-based adversarial examples. 

We evaluate 4 different autoencoder architectures, denoted as \textit{C}, \textit{A-1}, \textit{A-2}, and \textit{A-3}. C is the same architecture in the MagNet paper~\cite{meng2017magnet} for the CIFAR-10 dataset. Since the input size in our setting is much larger than that in CIFAR-10, we also evaluate 3 other architectures, A-1, A-2, and A-3, by adding 1, 2, 3 average pooling layers to C. From C, to A-3, the latent space dimension size decreases, which thus means more compression and lower input quality. All the autoencoder are trained with real-world images in KITTI dataset~\cite{kittidataset}(\msec{sec:evaluation}). 

\noindent {\bf Adversarial training (AT)~\cite{zhang2019towards}.} 
Since Apollo does not release the training dataset for its models, we can only evaluate this method on Y3 (YOLO v3). Since our attack needs to succeed for both camera and LiDAR, a secure camera model is still an effective defense strategy. We adapt our method to the state-of-the-art adversarial training-based method for camera-based object detection~\cite{zhang2019towards}. 
We follow their algorithm but change the attack in the training loop to ours. Since our attack is performed by adding an object instead of perturbing an existing one, an additional challenge is how to assign ground-truth bounding boxes and labels to our adversarial objects. To address this, we render benign object to the same position and use its detection results as ground-truth results for adversarial one. In AT, we only use bench object to perform experiment.

While adversarial training can be highly robust, it is known to be expensive and nearly intractable for large-scale problems~\cite{shafahi2019adversarial, wong2020fast}. In our case, this problem further exacerbates as the cost of our attack is higher than 2D digital-space attacks. Thus, we employ an acceleration method found in a recent work~\cite{wong2020fast} that allows a much smaller number of PGD steps (instead of a full optimization cycle) in each training iteration by randomly initializing the adversarial inputs. Specifically, we use a PGD with 2 step and randomly initialize the adversarial mesh during each training iteration. Besides, we train our model from a pre-trained Y3 model, which can converge much faster and also improve robustness~\cite{hendrycks2019using,chen2020adversarial}. 
We use the original Y3 training set COCO~\cite{coco}. We train the model for over 900 epoch, and the model converges after $\sim$83 epoch.

\noindent {\bf Augmenting training data (AUG)~\cite{szegedy:iclr:2014, goodfellow:fsgm, pei2017deepxplore}.} Prior works show that re-training the model with adversarial inputs mixed in the original training data can improve the model robustness~\cite{szegedy:iclr:2014, goodfellow:fsgm, pei2017deepxplore}. Same as for adversarial training, this method is only applied to Y3, and we use the same method as in adversarial training to generate the adversarial inputs and their ground-truth bounding boxes and labels. We use same COCO training dataset and the number of training epoch. In this case, the model converges at $\sim$48 epoch.

\end{document}